\documentclass[preprint,12pt]{elsarticle}

\usepackage[utf8]{inputenc}
\usepackage{graphicx}
\usepackage[T1]{fontenc}
\usepackage{amsmath}
\usepackage{amssymb}
\usepackage{float}
\usepackage{hyperref}
\usepackage{cleveref}
\usepackage{lmodern}
\usepackage{empheq}
\usepackage{braket}
\usepackage{xcolor}
\usepackage{cancel}
\usepackage{csquotes}
\usepackage{scalerel}
\usepackage{appendix}
\usepackage{afterpage}
\usepackage[mathscr]{eucal}
\usepackage[shortlabels]{enumitem}
\usepackage{import}
\usepackage{tikz}
\usetikzlibrary{shapes.geometric}
\usepackage{mathtools}
\usepackage{natbib}
\usepackage{orcidlink}
\usepackage{textgreek}

\hypersetup{
	colorlinks   = true, 
	urlcolor     = blue, 
	linkcolor    = blue, 
	citecolor   = blue 
}

\journal{Annals of Physics}

\begin{document}
	
	\begin{frontmatter}
		
		\title{Black Strings and String Clouds Embedded in Anisotropic Quintessence: Solutions for Scalar Particles and Implications}

		\author[inst1]{M.L. Deglmann \orcidlink{0000-0002-9737-5997} }
		\ead{m.l.deglmann@posgrad.ufsc.br}
		
		\affiliation[inst1]{organization={Departmento de Física, Universidade Federal de Santa Catarina (UFSC)},
			addressline={Campus Universitário Trindade}, 
			city={Florianópolis},
			postcode={88035-972}, 
			state={Santa Catarina},
			country={Brazil}}
		
		\author[inst1]{L.G. Barbosa \orcidlink{0009-0007-3468-3718}}
		\ead{leonardo.barbosa@posgrad.ufsc.br}
		
		\author[inst1]{C.C. Barros Jr. \orcidlink{0000-0003-2662-1844}}
		\ead{barros.celso@ufsc.br}

		\begin{abstract}
			We analyze the spacetime metric associated with a black string surrounded by a cloud of strings and an anisotropic fluid of quintessence in cylindrically symmetric AdS spacetime. We solve Einstein's equation to obtain the explicit form of the metric, investigate typical values for its parameters, and determine their role in the event horizon formation. Within our findings, we show that the intensity of the cloud of strings regulates the size of the event horizon and, when the cloud is absent, the horizon increases drastically for larger values of the quintessence's state parameter $\alpha_{\scaleto{Q}{4pt}}$. Additionally, the metric shows that, unless $\alpha_{\scaleto{Q}{4pt}}$ is close to its lower bound, the contribution from the quintessence fluid is only significant at large distances from the black string. Finally, to explore the quantum implications of this dark energy candidate, we use the confluent Heun function to solve the Klein-Gordon equation for a spin-0 particle near the event horizon. Our results indicate that the presence of quintessence alters the particle's radial wave function. This modification, in principle, could give rise to an observable that we termed as \enquote{dark phase}.
		\end{abstract}

		
	\end{frontmatter}
	
	\section{Introduction}
	
	In the late 1990s, the evidence for an accelerated expansion of the Universe, coming from astrophysical data of the luminosity distance of Type Ia supernovae \cite{Riess1998,Perlmutter1999}, dramatically changed our understanding of the Universe's history and evolution. The observed late-time accelerated cosmic expansion was attributed to the so-called dark energy (DE), a pervasive component of spacetime counteracting gravitational interaction \cite{COPELAND2006,Amendola2010}. In addition, other observational evidence for the existence of dark energy has emerged, including data from the age of the oldest stars, cosmic microwave background (CMB), baryon acoustic oscillations (BAO), and large-scale structure (LSS) (please, see \cite{COPELAND2006,Amendola2010,Steinhardt2003} and references therein).
	
	The simplest candidate for dark energy was Einstein's cosmological constant $\Lambda$, originally introduced to realize a static Universe \cite{Einstein1917}. 
	However, this interpretation faces some challenges, such as the fine-tuning and the coincidence problems \cite{Amendola2010}. Also, there are open questions such as, why dark energy should be constant in spacetime (if given by $\Lambda$). This question becomes critical if future data analysis from the Dark Energy Spectroscopic Instrument (DESI) confirms that dark energy evolves in time \cite{DESI32024,DESI42024,DESI62024}.
	
	To address these shortcomings, alternative models like quintessence have been proposed \cite{Caldwell1998}, where a canonical scalar field $\phi$ with a potential $V(\phi)$ offers a dynamical approach to dark energy, with a time-varying behavior and an energy density which did not need to be insignificant compared to matter and radiation in the early universe \cite{Amendola2010,Steinhardt2003,Caldwell1998}. The original construction from Caldwell, Dave, and Steinhardt \cite{Caldwell1998} was inspired by inflationary theory \cite{Guth:1980zm,Starobinsky:1980te,Linde:1981mu} and, in some subsequent models, inflation and dark energy were unified \cite{Hossain2015,Sa2020,Jimenez,Capozziello2006}.
	
	Another approach to studying quintessence involves defining an energy-momentum tensor $T^{\mu\nu}$, typically as an anisotropic fluid, with the desired properties of dark energy: positive-definite energy density and negative pressures, constrained by the state parameter $-1 < \omega_{\scaleto{Q}{4pt}} < -1/3$. One of the first works using this strategy was the one where Kiselev \cite{Kiselev2003} investigated the role of a quintessential fluid surrounding a spherically symmetric black hole (BH). Subsequent studies have explored various implications of this dark energy candidate for spacetime, thermodynamics, and surrounding particles \cite{Saadati2021,Toledo2018,Ghosh2016,Javed2024,Chaudhary2024,Zahid2024,Toshmatov2017}.
	
	Of particular interest is the work of Ali \textit{et al.} \cite{Ali2020}, who examined black strings, cylindrically symmetric Schwarzschild-like solutions in anti-de Sitter (AdS) spacetime \cite{Lemos1995,Lemos1996}, in the presence of quintessential fluid. One of the purposes of this work is to extend this configuration by adding a matter component analogous to a cloud of strings \cite{Letelier1979}. With this procedure, it will also be possible to investigate the effect of a mass distribution in this picture.
	
	As far as the quintessence hypothesis is considered, a fundamental question is how this kind of field may affect quantum systems and whether this effect is significant. This is another motivation for this work. Nowadays, there is an extensive investigation into how the structure of spacetime may affect the behavior of quantum systems. After the study of the one-electron atoms proposed by Parker \cite{Parker:1980hlc} many works have been made, such as the illustrative description of particles in cosmic strings backgrounds \cite{Santos:2017eef,Santos:2016omw,Vitoria:2018its}, Kerr black holes 
	\cite{chandra} in the Schwarzschild spacetime \cite{elizalde_1987} and others \cite{Sedaghatnia:2019xqb,Guvendi:2022uvz,Vitoria:2018mun,Maniccia:2023cgv}.
	The Casimir effect \cite{Santos:2018jba}, and bosons in the Hartle-Thorne spacetime \cite{Pinho:2023nfw} are further interesting examples that can be pointed out. Another important kind of system, which is relevant for many branches of physics, is the quantum oscillator. It has been studied in different formulations \cite{Ahmed:2022tca,Ahmed:2023blw,Santos:2019izx,Yang:2021zxo,Soares:2021uep,Rouabhia:2023tcl} for many kinds of background. All of these works provided many interesting results and are responsible for the current progress on this question. 
	
	This research explores the behavior of quantum systems, specifically spin-$0$ particles, in a curved spacetime influenced by dark energy and its relative contribution when compared to the mass and cosmological constant components that determine the geometry of this spacetime. To achieve this, we consider a cosmological model comprised of a black string, quintessence, and string clouds, embedded in an anti-de Sitter spacetime with a negative cosmological constant, $\Lambda = -3/l^{2}$, where $l$ stands for the AdS radius. Our primary focus is on understanding how quintessence, as a representative of dark energy, affects the spacetime metric and subsequently the solutions of the Klein-Gordon (KG) equation. By employing the confluent Heun equation \cite{Ronveaux1995,Olver1997,Olver2010,Hortacsu2018,ChebTerrab2004}, we seek analytical solutions for the KG equation in this background.

	This paper is organized as follows: in \cref{sec-metric-definition} we choose the spacetime metric, obtain the Einstein tensor, and determine the appropriate contributions of the energy-momentum tensor from quintessence and the clouds of strings. In \cref{sec-Einstein_Equations}, we solve Einstein equations to determine the metric function $A(\rho)$ and give the explicit result for the quintessence energy density and pressures. Then, in \cref{sec-Estimates}, we investigate admissible values for the parameters occurring in the metric, written in terms of SI units. Next, in \cref{sec-Event-Horizon}, we investigate the event horizon formation and give analytic results related to specific values of quintessence's parameter of state. Subsequently, in \cref{sec-KG-equation} we solve the Klein-Gordon equation for a spin-$0$ particle in this background, near the event horizon, and define an observable which we termed as \enquote{dark phase}. Finally, we summarize our key findings in \cref{sec-Conclusion}. \ref{ap-units} contains a dimensional analysis of our parameters in terms of SI units. \ref{app-Heun-equation} presents the most important results regarding the confluent Heun equation.

	\section{Metric definition and Einstein tensor}\label{sec-metric-definition}
	
	To describe a spacetime whose structure is determined by a black string with a cloud of strings, an anisotropic Kiselev fluid, and a cosmological constant, let us consider the following interval
	\begin{equation}
		ds^{2} = A(\rho)\,c^{2}dt^{2} - 	\frac{d\rho^{2}}{A(\rho)} - \rho^{2}d\varphi^{2} - \frac{\rho^{2}}{l^{2}}\,dz^{2}\,,\label{metric}
	\end{equation}
	associated with an ansatz similar to \citep{Ali2020}. The function $A(\rho)\in\mathbb{R}$, while $\rho,\,\varphi\,\text{and }z$ are the standard cylindrical coordinates. Additionally, $l^{2} = -3/\Lambda$, with $\Lambda$ being the cosmological constant.
	
	We take the general Einstein tensor to be
	\begin{equation}
		G_{\mu\nu} = R_{\mu\nu} - \frac{1}{2}g_{\mu\nu}\,R\,,
	\end{equation}
	where $R_{\mu\nu}$ is the Ricci tensor while $R$ is the Ricci scalar. Then, for the metric tensor associated with \cref{metric}, the (cylindrical) components $G_{\mu}^{\ \nu}$ of the Einstein tensor become
	\begin{align}
		G_{t}^{\ t} = G_{\rho}^{\ \rho}
		&= \frac{1}{\rho}\,\frac{dA}{d\rho} + \frac{A}{\rho^{2}}\,,\label{Einstein_t_ut}\\[5pt]
		G_{\varphi}^{\ \varphi} = G_{z}^{\ z}
		&= \frac{1}{2}\,\frac{d^{2}A}{d\rho^{2}} + \frac{1}{\rho}\,\frac{dA}{d\rho}\,.\label{Einstein_phi_uphi}
	\end{align}
	Since the Einstein equations are given by
	\begin{equation}
		G_{\mu}^{\ \nu} + \Lambda\,\delta_{\mu}^{\ \nu} = \frac{8\pi G}{c^{4}}\,T_{\mu}^{\ \nu}\,,\label{general-Einstein-equations}
	\end{equation}
	we can construct an appropriate ansatz for the energy-momentum tensor. Moreover, please note that we have chosen to work in the SI units convention. 
	
	The Einstein tensor, in the considered metric, is given by \cref{Einstein_t_ut,Einstein_phi_uphi}, and together with the Einstein equations \eqref{general-Einstein-equations}, implies that an energy-momentum tensor compatible with this line element must be diagonal and symmetric. This means that the off-diagonal components of the energy-momentum tensor must be zero, and the diagonal components must satisfy
	\begin{equation}
		\begin{aligned}	
			T_{t}^{\ t} &= T_{\rho}^{\ \rho}\,,\\
			T_{\varphi}^{\ \varphi} &= T_{z}^{\ z}\,.\\
		\end{aligned} \label{General-T-munu}
	\end{equation}
	In other words, the Einstein equations impose a symmetry requirement on the energy-momentum tensor.
	
	\subsection{The energy-momentum tensor}
	
	We propose a general ansatz for the energy-momentum tensor, according to \cref{General-T-munu}, given by 
	\begin{equation}
		\begin{aligned}
			T_{t}^{\ t} &= T_{\rho}^{\ \rho} = \sigma(\rho)\,,\\
			T_{\varphi}^{\ \varphi} &= T_{z}^{\ z} = \zeta(\rho)\,,
		\end{aligned}\label{cyl-sym-ansatz}
	\end{equation}
	with functions $\sigma(\rho)\,,\,\zeta(\rho)\,\in\,\mathbb{R}$ depending only on $\rho$ for cylindrical symmetry.
	
	The conservation law for the energy-momentum tensor, derived from
	\begin{equation*}
		T_{\alpha\,\,\, ;\beta}^{\,\,\beta} = \partial_{\beta}\,T_{\alpha}^{\,\,\beta} - \Gamma^{\lambda}_{\,\,\alpha\beta}\,T_{\lambda}^{\,\,\beta} + \Gamma^{\beta}_{\,\,\lambda\beta}\,T_{\alpha}^{\,\,\lambda} = 0\,,\label{Conservation-Law-Formula}
	\end{equation*}
	where $\Gamma^{\lambda}_{\,\,\alpha\beta}$ are the Christoffel symbols, imposes a constraint on $\sigma(\rho)$ and $\zeta(\rho)$, given by
	\begin{equation}
		\frac{d}{d\rho}\,\sigma(\rho) + \frac{2}{\rho}\left[\sigma(\rho) - \zeta(\rho)\right] = 0\,.\label{Eq_Conservation}  
	\end{equation}
	In principle, if we can determine the relationship between $\sigma(\rho)$ and $\zeta(\rho)$, we can solve this equation and obtain the explicit form of the energy-momentum tensor $T_{\mu}^{\ \nu}$ for that system, and the result satisfies the conservation law.
	
	The condition \eqref{Eq_Conservation} is quite general and allows many possibilities for the forms of these functions. 
	A particular case arises when $\zeta(\rho)= k\,\sigma(\rho)$, where $k \in \mathbb{R}$ is a constant. In this case, the conservation law simplifies to
	\begin{equation}
		\frac{d}{d\rho}\,\sigma(\rho) = \frac{2(k-1)}{\rho}\,\sigma(\rho)\,,
	\end{equation}
	leading to the solution
	\begin{equation}
		\sigma(\rho) = \sigma_{0}\,\rho^{2(k-1)}\,,\label{energy-momentum-cases}
	\end{equation}
	where $\sigma_0$ is a constant.
	We present three specific cases for the energy-momentum tensor, as shown in \cref{table:energy-momentum-cases}.
	\begin{table}
		\begin{center}
			\begin{tabular}{|c|c|c|}
				\hline
				$k$ & $T_{t}^{\ t} = T_{\rho}^{\ \rho}$ &	$T_{\varphi}^{\ \varphi} = T_{z}^{\ z}$\\[5pt]
				\hline \hline
				$0$ & $\sigma_{0}/\rho^{2}$ & $0$\\[5pt]
				$1$ & $\sigma_{0}$ & $\sigma_{0}$\\[5pt]
				$\alpha_{\scaleto{Q}{4pt}}$ & $\sigma_{0}\,\rho^{2(\alpha_{\scaleto{Q}{3pt}}-1)}$ & $\alpha_{\scaleto{Q}{4pt}}\,\sigma(\rho)$\\
				\hline
			\end{tabular}
			\caption{Special cases of the energy-momentum tensor given by \cref{energy-momentum-cases}.}
			\label{table:energy-momentum-cases}
		\end{center}
	\end{table}
	To explore this system, in this work, we will focus our attention on the first and third cases described by \cref{table:energy-momentum-cases}, where $\alpha_{\scaleto{Q}{4pt}}\in\mathbb{R}$. Due to the conservation of the energy-momentum tensor and the linearity of the covariant derivative, the sum of these particular cases is also conserved.
	
	\subsubsection{The contribution from quintessence and clouds of strings}
	
	Following the work of \cite{Ali2020}, we describe the energy-momentum tensor for the quintessence fluid as
	\begin{equation}
		\begin{aligned}
			T_{t}^{\ t} &= T_{\rho}^{\ \rho} = \rho_{\scaleto{Q}{4pt}}(\rho)\,,\\
			T_{\varphi}^{\ \varphi} &= T_{z}^{\ z} = \alpha_{\scaleto{Q}{4pt}}\,\rho_{\scaleto{Q}{4pt}}(\rho)\,,
		\end{aligned}\label{Quintessence_EM_Tensor}
	\end{equation}
	where
	\begin{equation}
		\alpha_{\scaleto{Q}{4pt}} \coloneq -\frac{1}{2}\left(3\, \omega_{\scaleto{Q}{4pt}} + 1\right)\,, \label{eq:Alpha_Q_Definition}
	\end{equation}
	with $-1<\omega_{\scaleto{Q}{4pt}}<-1/3$. Therefore, we consider $\alpha_{\scaleto{Q}{4pt}}$ to be a constant (time-independent) parameter, taking values in the interval $0<\alpha_{\scaleto{Q}{4pt}}<1$. This form is consistent with the third case in \cref{table:energy-momentum-cases}, ensuring that the energy-momentum tensor obeys the conservation law.
	
	The non-trivial components of this energy-momentum tensor should be interpreted as
	\begin{equation}
		\begin{aligned}
			T_{t}^{\ t} &= \rho_{\scaleto{Q}{4pt}}\,,\\
			T_{\rho}^{\ \rho} &= -p_{\rho}\,,\\
			T_{\varphi}^{\ \varphi} &= -p_{\varphi}\,,\\
			T_{z}^{\ z} &= -p_{z}\,,
		\end{aligned}\label{fluid_T_mu_nu}
	\end{equation}
	implying that the fluid has energy density $\rho_{\scaleto{Q}{4pt}}$ and the pressure is given by the relations $p_{\rho} = -\rho_{\scaleto{Q}{4pt}}$, $p_{\varphi} = p_{z} = -\alpha_{\scaleto{Q}{4pt}}\,\rho_{\scaleto{Q}{4pt}}$.
	This means that the fluid is anisotropic unless $\alpha_{\scaleto{Q}{4pt}}=1$ occurs (in the limiting case of $\omega_{\scaleto{Q}{4pt}} = -1$).
	
	In addition to the quintessence fluid, we introduce an extra matter component, representing a distribution of matter, having the form of \cref{energy-momentum-cases} with $k=0$. It is analogous to the so-called \emph{"clouds of strings"} \cite{Letelier1979} and is described by
	\begin{equation}
		\begin{aligned}
			T_{t}^{\ t} &= T_{\rho}^{\ \rho} = \frac{a}{\rho^{2}}\,,\\
			T_{\varphi}^{\ \varphi} &= T_{z}^{\ z} = 0\,,
		\end{aligned}\label{T-munu-Cloud-of-strings}
	\end{equation}
	where $a \in \mathbb{R}^{+}$ is a positive constant. This matter component possesses only energy density and radial pressure, distinguishing it from the quintessence fluid.

	\section{The Einstein equations} \label{sec-Einstein_Equations}
	
	To solve the Einstein equations, we combine the components of the Einstein tensor, given by \cref{Einstein_t_ut,Einstein_phi_uphi}, with the energy-momentum tensor from quintessence, defined by \cref{Quintessence_EM_Tensor}, and the one from clouds of strings, described by \cref{T-munu-Cloud-of-strings}. This yields the following set of equations:
	\begin{align}
		\frac{1}{\rho}\,\frac{dA}{d\rho} + \frac{A}{\rho^{2}} +\Lambda
		&= \frac{8\pi G}{c^{4}}\left(\rho_{\scaleto{Q}{4pt}} + \frac{a}{\rho^{2}}\right)\,,\label{Einstein_Eq_1}\\
		\frac{1}{2}\,\frac{d^{2}A}{d\rho^{2}} + \frac{1}{\rho}\,\frac{dA}{d\rho} + \Lambda
		&= \frac{8\pi G}{c^{4}} \alpha_{\scaleto{Q}{4pt}}\,\rho_{\scaleto{Q}{4pt}}\,.\label{Einstein_Eq_2}
	\end{align}
	To find the solution for $A(\rho)$, we multiply \cref{Einstein_Eq_1} by $\alpha_{\scaleto{Q}{4pt}}$ and subtract it from \cref{Einstein_Eq_2}. Next, we multiply the resultant equation by $2\rho^{2}$ so that it becomes a nonhomogeneous Cauchy-Euler equation for $A(\rho)$
	\begin{equation}
		\rho^{2}\,\frac{d^{2}A}{d\rho^{2}} + 2(1-\alpha_{\scaleto{Q}{4pt}})\rho\,\frac{dA}{d\rho} - 2\, \alpha_{\scaleto{Q}{4pt}}\,A 
		= -\frac{16\pi G}{c^{4}}\,a\,\alpha_{\scaleto{Q}{4pt}}\ ,\label{Eq_for_A}
	\end{equation}
	implying the homogeneous solution $A_{h}(\rho)$ is given by
	\begin{equation*}
		A_{h}(\rho) = -\frac{2G}{c^{2}}\frac{m_{\scaleto{S}{4pt}}}{\rho} + N_{\scaleto{Q}{4pt}}\,\rho^{2\,\alpha_{\scaleto{Q}{3pt}}}\,,\label{A_h}
	\end{equation*}
	where $m_{\scaleto{S}{4pt}}$ and $N_{\scaleto{Q}{4pt}}$ are positive real integration constants. Note that we set the first integration constant to be $-2Gm_{\scaleto{S}{4pt}}/c^{2}$ so that $m_{\scaleto{S}{4pt}}$ can be interpreted as a Schwarzschild-like mass contribution, in SI units.
	
	To find the particular solution $A_{p}(\rho)$, we assume the form $A_{p}(\rho) = c_{1} + C_{2}\rho^{2}$ and, after plugging it into \cref{Eq_for_A}, we find $c_{1} = 8\pi G/a c^{4}$ and $c_{2} = 1/l^{2}$. Therefore, the particular solution is $A_{p}(\rho) = 8\pi G/{a c^{4}} + \rho^{2}/l^{2}$ and the solution for $A(\rho)$ reads	
	\begin{equation}
		A(\rho) = \overline{a} + \frac{\rho^{2}}{l^{2}} - \frac{\rho_{\scaleto{S}{4pt}}}{\rho} + N_{\scaleto{Q}{4pt}}\,\rho^{2\alpha_{\scaleto{Q}{3pt}}}\,,\label{Function_A(rho)}
	\end{equation}
	where
	\begin{align}
		\overline{a} &\coloneq \frac{8\pi G}{c^{4}}\,a\,,\label{a-bar-def}\\
		\rho_{\scaleto{S}{4pt}} &\coloneq \frac{2G}{c^{2}}\, m_{\scaleto{S}{4pt}}\,, \label{Schwarzschild-like-def}
	\end{align}
	are the dimensionless cloud parameter and the Schwarzschild-like radius, respectively. These results allow us to determine the quintessence energy density and show that it is independent of the cosmological constant. Thinking in terms of the physics of this system, it is interesting to investigate the possible values for these physical parameters, a task that will be carried out in section \ref{sec-Estimates}.

	\subsection{Quintessence energy density and pressure}\label{energy-density-pressure}
	
	Since we have determined $A(\rho)$, given by \cref{Function_A(rho)}, we use \cref{Einstein_Eq_1} to obtain the quintessence energy density $\rho_{\scaleto{Q}{4pt}}$, resulting in
	\begin{equation}
		\rho_{\scaleto{Q}{4pt}} = \left(2\, \alpha_{\scaleto{Q}{4pt}} + 1\right)\frac{N_{\scaleto{Q}{4pt}}}{8\pi G/c^{4}}\,\rho^{2\, \alpha_{\scaleto{Q}{3pt}}-2}\,.\label{Energy_Density}
	\end{equation}
	Given that the energy density must be positive, we can see that the condition $N_{\scaleto{Q}{4pt}}\geq 0$ is compatible with the range of values $1<2\, \alpha_{\scaleto{Q}{4pt}} +1<3$. 
	
	The results for tangent and longitudinal pressures, $p_{\varphi}$ and $p_{z}$, follow from \cref{Quintessence_EM_Tensor} and are given by
	\begin{equation}
		p_{\scaleto{\varphi}{5pt}} = p_{\scaleto{z}{4pt}} = -\frac{\alpha_{\scaleto{Q}{4pt}}\left(2\, \alpha_{\scaleto{Q}{4pt}} + 1\right)}{8\pi G}\,c^{4}\,N_{\scaleto{Q}{4pt}}\,\rho^{2\, \alpha_{\scaleto{Q}{3pt}}-2}\,,\label{Pressures}
	\end{equation}
	and it is noteworthy that $p_{\varphi}$ and $p_{z}$ are indeed negative, as we expected. Furthermore, it is intriguing to note that the energy density, $\rho_{\scaleto{Q}{4pt}}$, and its corresponding pressures remain unchanged regardless of the value of the cosmological constant, $\Lambda$.
	
	In the formulation presented in this section, many parameters and effects have been considered. To obtain a better understanding of this system, thinking in terms of its physical content, the next section will present estimates of their values for different astrophysical objects, results that will be useful when considering quantum particles inside this background.

	\section{Estimates for the physical values of parameters}\label{sec-Estimates}
	
	In the formulation of the metric, presented in \cref{sec-metric-definition,sec-Einstein_Equations}, some parameters have been considered. Depending on the values of these parameters, different physical scenarios may appear. So, to provide a better understanding of these systems, we will study these parameters and relate them with typical astrophysical objects in this section.
	
	To estimate the possible values for $a$, $\rho_{\scaleto{S}{4pt}}$, and $N_{\scaleto{Q}{4pt}}$, as well as to determine their influence on $A(\rho)$, it is helpful to establish the dimensions of these quantities. This analysis is performed in \ref{ap-units} using SI units. In what follows, we use the insight from this dimensional analysis to find an admissible range of values for these parameters. We must remark that, in this section, we are interested in estimating the orders of magnitude for the parameters, not their exact values.

	\subsection{String clouds parameter}
	
	There are two ways to estimate reasonable values for the parameter $a$ describing the clouds of strings. 
	
	The first one is to choose a range of objects with cylindrical symmetry such as, for instance, disks of spiral galaxies, and use them to determine the energy per length in such objects. A naive, but still valid, estimate would be $a \geq \text{disk mass} \times c^{2}/{\text{disk thickness}}$.
	Considering a mass range of $10^{39} - 10^{42}$ kg and thicknesses varying from $10^{20} - 10^{21}$ m, we obtain that $a$ would range from $10^{36} - 10^{38}$ J/m$^{3}$. However, according to the definition \eqref{a-bar-def}, what matters in $A(\rho)$ is the dimensionless cloud parameter $\overline{a}$, which according to the aforementioned estimate, would take values in the interval $10^{-7} - 10^{-5}$. Note that we use the $2022$ \textsc{CODATA} recommended values of the fundamental physical constants \cite{CODATA2022} for $G$ and $c$, in SI units.
	
	Another way to estimate the parameters from a cylindrical spacetime using data from our Universe is to transpose the mass from spherical objects into an equivalent cylindrical shape. To preserve volume, we choose a cylinder with a radius $r_{c}$ and height $2\,r_{c}$, where:
	\begin{equation}
		r_{c} = \left(\frac{2}{3}\right)^{1/3} r_{s}\,.\label{cylinder-vs-sphere}
	\end{equation}
	Here, $r_{s}$ is the radius of the original spherical object. Using this approach, we can estimate the equivalent parameter $a$ for spherical objects as:
	\begin{equation}
		a \geq \frac{m_{\scaleto{obj.}{6pt}}}{(16/3)^{1/3} \,r_{s}}\, c^{2}\,.
	\end{equation}
	This allows us to estimate $a$ for various celestial bodies, such as stars, planets, and nebulae. As an example, we present \cref{table:Dimensionless_a_bar} to show that typical values of $\overline{a} = a\,(8\pi G/c^{4})$ range from $10^{-12} - 10^{1}$.
	
	\begin{table}
		\begin{center}	
			\begin{tabular}{| c | c | c | c |}
				\hline
				Stellar Object & Mass (kg) & Radius (m) & $\overline{a}$\\
				\hline \hline
				Crab Nebula & $6.0 \times 10^{30}$ & $5.2 \times 10^{16}$ & $1.2 \times 10^{-12}$\\
				Earth & $6.0\times 10^{24}$ & $6.3\times 10^{6}$ & $1 .0\times 10^{-8}$\\
				Betelgeuse & $3.8\times 10^{31}$ & $5.3 \times 10^{11}$ & $7.6 \times 10^{-7}$\\
				Rigel & $4.2\times 10^{31}$ & $5.2 \times 10^{10}$ & $8.6 \times 10^{-6}$\\
				Sun & $2.0\times 10^{30}$ & $7.0\times 10^{8}$ & $3.0\times 10^{-5}$\\
				Proxima Centauri & $2.4\times 10^{29}$ & $1.1 \times 10^{8}$ & $2.3 \times 10^{-5}$\\
				Messier 87 & $6.0 \times 10^{42}$ & $1.4 \times 10^{21}$ & $4.6 \times 10^{-5}$\\
				RMC136a1 & $3.9 \times 10^{32}$ & $3.0 \times 10^{10}$ & $1.4 \times 10^{-4}$\\
				White Dwarfs & $1.2\times 10^{30}$ & $7.0 \times 10^{6}$ & $1.8\times 10^{-3}$\\
				Crab Pulsar & $2.8\times 10^{30}$ & $1.0\times 10^{4}$ & $3.0 \times 10^{0}$\\
				\hline
			\end{tabular}
		\end{center}\caption{Estimated dimensionless parameter $\overline{a}$ for various stellar objects. Note: Data is approximate and sourced from \cite{CrabNebulaMass1985,CrabNebulaRadius2008,ParticleDataGroup2024,Betelgeuse2020,RigelMass2013,RigelRadius2017,Proxima2017,Messier1997,R136Mass2022,R136Radius2022,WhiteDwarfsMass2007,WhiteDwarfsRadius1979}.} 
		\label{table:Dimensionless_a_bar} 
	\end{table}

	\subsection{Schwarzschild-like radius}\label{Estimates_R_s}
	
	To explore reasonable values for the Schwarzschild-like radius $\rho_{\scaleto{S}{4pt}}$, let us recall it is defined as 	$\rho_{\scaleto{S}{4pt}} = 2\,G m/c^{2}$. We must remember that $\rho_{\scaleto{S}{4pt}}$ is not the radius of the event horizon, it is a parameter with a mathematical definition similar to the usual Schwarzschild radius.
	Using the known value of $G/c^2 \,(7.416 \times 10^{-28}\text{ kg}^{-1}\text{ m})$ \cite{ParticleDataGroup2024}, we can calculate $\rho_{\scaleto{S}{4pt}}$ associated with typical stellar objects, in analogy with the previous section. Considering masses in the range of $10^{25} - 10^{42}$ kg, for instance, we would obtain Schwarzschild-like radii in the range of $10^{-2} - 10^{15}$ m, respectively.
	
	Finally, we note that the influence of $\overline{a}$ on $A(\rho)$ is only significant for large values of $\rho$. This happens because the impact of $\rho_{\scaleto{S}{4pt}}$ gradually decreases as $\rho$ increases.
	
	\subsection{Quintessence parameter}
	
	To estimate a reasonable value for $N_{\scaleto{Q}{4pt}}$, we must recall the result from the energy density from quintessence, given by \eqref{Energy_Density}:
	\begin{equation*}
		\rho_{\scaleto{Q}{4pt}} = \left(2\alpha_{\scaleto{Q}{4pt}} + 1\right) \overline{N}_{\scaleto{Q}{4pt}}\,\rho^{2\alpha_{\scaleto{Q}{4pt}}-2}\,,
	\end{equation*}
	with
	\begin{equation*}
		\overline{N}_{\scaleto{Q}{4pt}} \coloneqq \frac{N_{\scaleto{Q}{4pt}}}{8\pi G/c^{4}}\,.
	\end{equation*}
	Taking into consideration that Dark Energy's energy density is of the order of $\rho_{\scaleto{DE}{4pt}} \approx 6 \times 10^{-10}$ J/m$^{3}$ \cite{ParticleDataGroup2024,PlanckCosmology2014}, we can obtain an estimate for the possible values of $\overline{N}_{\scaleto{Q}{4pt}}$ in the limiting case of $\alpha_{\scaleto{Q}{4pt}} \to 1$, resulting in
	\begin{equation}
		\rho_{\scaleto{Q}{4pt}}^{(\alpha_{\scaleto{Q}{4pt}}\to1)} = 3\, \overline{N}_{\scaleto{Q}{4pt}} \approx 6 \times 10^{-10} \text{ J/m}^{3}\,,
	\end{equation}
	so that $\overline{N}_{\scaleto{Q}{4pt}} \approx 2 \times 10^{-10}$ J/m$^{3}$ when $\alpha_{\scaleto{Q}{4pt}} \to 1$. This result means that $N_{\scaleto{Q}{4pt}}$ is of the order
	\begin{equation}
		N_{\scaleto{Q}{4pt}} \approx 4.1 \times 10^{-53}\text{ m}^{-2}\,,
	\end{equation}
	which is impressively small. Note that result might change if $\alpha_{\scaleto{Q}{4pt}}$ varies.
	
\begin{figure}[ht!]
		\begin{center} 
			\includegraphics[width=0.65\textwidth]{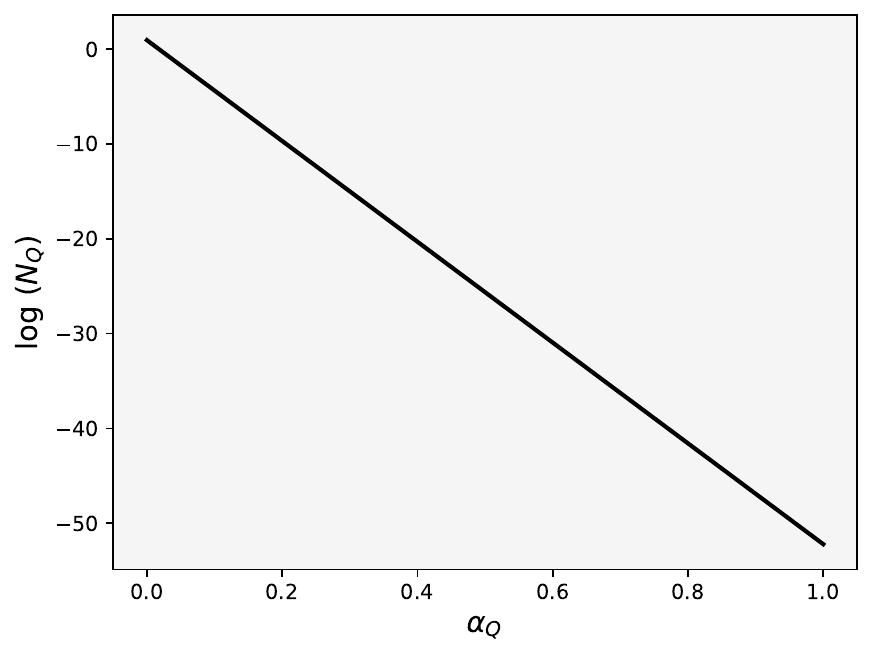}
			\caption{Logarithmic plot of the quintessence parameter $N_{\scaleto{Q}{4pt}}$ as a function of $\alpha_{\scaleto{Q}{4pt}}$. According to \cref{Estimate-N_Q}, there is an exponential decrease in $N_{\scaleto{Q}{4pt}}$ when the state parameter $\alpha_{\scaleto{Q}{4pt}}$ increases. The vertical axis values are calculated assuming an observable universe with a radius of $\rho_{\scaleto{obs.}{4pt}} = 3.8 \times 10^{26}$ m, ensuring that this cylindrical model has the same observable volume as our actual universe.}
			\label{N_Q}
		\end{center}
	\end{figure}

	To estimate a reasonable order of magnitude for $N_{\scaleto{Q}{4pt}}$, for different values of $\alpha_{\scaleto{Q}{4pt}}$, we need an alternative approach. Let us start by calculating the total quintessence energy contribution within the observable universe. 
	First, let us recall that the current estimate for the radius of the observable Universe, according to $\Lambda$CDM cosmology, is roughly $r_{\scaleto{obs.}{4pt}} \approx 4.4 \times 10^{26}$ m \cite{ParticleDataGroup2024,PlanckCosmology2014}, assuming a spherical shape. By calculating the equivalent radius of a cylinder with the same volume, employing \cref{cylinder-vs-sphere}, we achieve an estimate for a cylindrical Universe, where
	\begin{equation}
		\rho_{obs.} \approx \left(2/3\right)^{1/3}\,4.4 \times 10^{26}\text{ m} \approx 3.8 \times 10^{26}\text{ m}\,.\label{rho-obs}
	\end{equation}
	Hence, we can integrate the energy density \eqref{Energy_Density}, i.e. $\rho_{\scaleto{Q}{4pt}}(\rho)$, from $\rho = 0$ to $\rho_{obs.}$ in order to show that
	\begin{equation*}
		\begin{aligned}	
			E_{\scaleto{Q}{4pt}} 
			= \left(2\alpha_{\scaleto{Q}{4pt}} + 1\right) \overline{N}_{\scaleto{Q}{4pt}}\,\int_{S_{1}^{obs.}\times \mathbb{R}}\,\rho^{2\alpha_{\scaleto{Q}{4pt}}-2}\,dV  = \frac{2\pi\,\overline{N}_{\scaleto{Q}{4pt}}}{|l|}\,\Delta z\,\rho_{obs.}^{2\alpha_{\scaleto{Q}{4pt}} + 1}\,.
		\end{aligned}
	\end{equation*}
	To maintain the same volume as the spherical universe, we set the cylinder's height to $\Delta z = 2 \rho_{\scaleto{obs.}{4pt}}$. This leads to
	\begin{equation*}
		E_{\scaleto{Q}{4pt}} = \frac{4\pi\,\overline{N}_{\scaleto{Q}{4pt}}}{|l|}\,\rho_{obs.}^{2\alpha_{\scaleto{Q}{4pt}} + 2}\,.
	\end{equation*}
	Assuming that the radius of anti-de Sitter space, $l$, is such that $|l| \gtrsim \rho_{obs.}$, we can estimate the final contribution of quintessence energy, $E_{\scaleto{Q}{4pt}}$, to be
	\begin{equation}
		E_{\scaleto{Q}{4pt}} \approx 4\pi \overline{N}_{\scaleto{Q}{4pt}}\,\rho_{obs.}^{2\alpha_{\scaleto{Q}{4pt}} + 1}\,.\label{DE-com-alpha}
	\end{equation}
	To compare the quintessence energy contribution ($E_{\scaleto{Q}{4pt}}$) with the total dark energy in the $\Lambda$CDM universe ($E_{\scaleto{DE}{4pt}}$), we first need to estimate $E_{\scaleto{DE}{4pt}}$:
	\begin{equation}
		E_{\scaleto{DE}{4pt}} \approx \frac{4\pi r_{\scaleto{obs.}{4pt}}^{3}}{3}\,\rho_{\scaleto{DE}{4pt}} \approx 2 \times 10^{71}\text{ kg m}^{2}\text{ s}^{-2}\,.\label{DE-Estimate}
	\end{equation}
	Therefore, comparing the results \eqref{DE-com-alpha} and \eqref{DE-Estimate}, with $\rho_{obs.}$ given by \eqref{rho-obs}, we obtain that
	\begin{equation}
		N_{\scaleto{Q}{4pt}} = \frac{2\,G}{c^{4}}\,\frac{E_{\scaleto{DE}{4pt}}}{\rho_{obs.}^{2\alpha_{\scaleto{Q}{4pt}} + 1}} = \frac{3.4 \times 10^{27}}{\rho_{obs.}^{2\alpha_{\scaleto{Q}{4pt}} + 1}}\,.\label{Estimate-N_Q}
	\end{equation}
	We show numerical estimates of $N_{\scaleto{Q}{4pt}}$ for a large universe with $\rho_{\scaleto{obs.}{4pt}}$ given by \eqref{rho-obs} in \cref{N_Q}.
	
	It can be readily shown that as $\alpha_{\scaleto{Q}{4pt}}$ increases, $N_{\scaleto{Q}{4pt}}$ decreases many orders of magnitude rapidly. This happens because the total quintessence energy $E_{\scaleto{Q}{4pt}}$ is an increasing function of $\rho^{2\alpha_{\scaleto{Q}{4pt}}}$ when $\alpha_{\scaleto{Q}{4pt}} \neq 0$. Hence, as $\alpha_{\scaleto{Q}{4pt}}$ increases, $\lim_{\alpha_{\scaleto{Q}{4pt}}\to 1} N_{\scaleto{Q}{4pt}} = 0$,
	so that $N_{\scaleto{Q}{4pt}}$ balances the expected value for $E_{\scaleto{Q}{4pt}}$.
	
	With the considerations above, the contribution from quintessence in $A(\rho)$ is given by
	\begin{equation}
		N_{\scaleto{Q}{4pt}}\,\rho^{2\alpha_{\scaleto{Q}{4pt}}} = \frac{2\,G}{c^{4}}\,\frac{E_{\scaleto{DE}{4pt}}}{\rho_{\scaleto{obs.}{4pt}}}\,\left(\frac{\rho}{\rho_{\scaleto{obs.}{4pt}}}\right)^{2\alpha_{\scaleto{Q}{4pt}}}\,,
	\end{equation}  
	which becomes
	\begin{equation}
		N_{\scaleto{Q}{4pt}}\,\rho^{2\alpha_{\scaleto{Q}{4pt}}} = \frac{3.4 \times 10^{27}}{\rho_{\scaleto{obs.}{4pt}}}\,\left(\frac{\rho}{\rho_{\scaleto{obs.}{4pt}}}\right)^{2\alpha_{\scaleto{Q}{4pt}}} =
		8.95\,\left(\frac{\rho}{\rho_{\scaleto{obs.}{4pt}}}\right)^{2\alpha_{\scaleto{Q}{4pt}}}\,.\label{contribution-N_Q_Rho_Alpha_Q}
	\end{equation}
	where the  last equality is true for $\rho_{\scaleto{obs.}{4pt}}$ given by \eqref{rho-obs}.
	
\begin{figure}[ht!]
		\centering
		\includegraphics[width=0.5\textwidth]{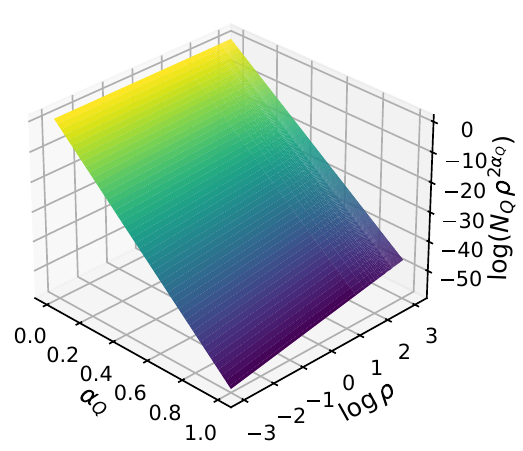}
		\caption{The contribution of the quintessence term, $N_{\scaleto{Q}{4pt}}\,{\rho}^{2\alpha_{\scaleto{Q}{4pt}}}$, to the spacetime metric function $A(\rho)$ was analyzed. Using an observable universe radius of 
			$l = \rho_{\scaleto{obs.}{4pt}} = 3.8 \times 10^{26}$ m, we found that the quintessence term's significance is most pronounced for small values of the state parameter $\alpha_{\scaleto{Q}{4pt}}$, approaching zero. Conversely, as $\alpha_{\scaleto{Q}{4pt}}$ increases towards $\alpha_{\scaleto{Q}{4pt}} = 1$, the term becomes relevant only at very large radii, comparable to the radius of the observable universe $\rho_{\scaleto{S}{4pt}}$.}
		\label{fig_Quintessence_Contribution}
	\end{figure}

	Our analysis indicates that the quintessence contribution to the spacetime metric function $A(\rho)$ is significant only for $\alpha_{\scaleto{Q}{4pt}} \ll 1$ or when $\rho$ is large enough compared to $\rho_{\scaleto{obs.}{4pt}}$. In \cref{fig_Quintessence_Contribution}, we show the behavior of the term $N_{\scaleto{Q}{4pt}}\,{\rho}^{2\alpha_{\scaleto{Q}{4pt}}}$ for a range of values of $\rho$ and for the whole interval $0<\alpha_{\scaleto{Q}{4pt}}<1$.
	
	\subsection{Analysis of the metric profile function}
	
	After estimating plausible values for the parameters $\overline{a}$, $\rho_{\scaleto{S}{4pt}}$, and $N_{\scaleto{Q}{4pt}}$, we can explore their role in $A(\rho)$, determined by \cref{Function_A(rho)}.
	
\begin{figure}[ht]
		\centering
		\begin{minipage}[t]{0.48\textwidth}
			\centering
			\includegraphics[width=\textwidth]{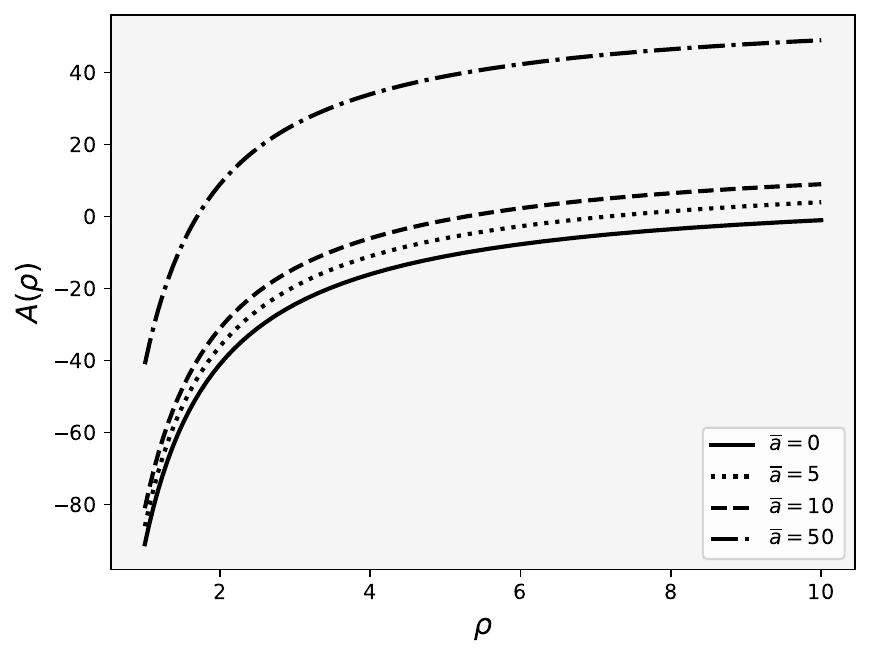}
			\caption{Plot of $A(\rho)$ for different values of $\overline{a}$. In this figure, $\alpha_{\scaleto{Q}{4pt}} = 0$, $\rho_{\scaleto{S}{4pt}} = 10^{2}$ m, $l = \rho_{\scaleto{obs.}{4pt}} = 3.8 \times 10^{26}$ m, and $N_{\scaleto{Q}{4pt}} = 8.95$.}
			\label{figure-a_bar_variando}
		\end{minipage}
		\hfill
		\begin{minipage}[t]{0.48\textwidth}
			\centering
			\includegraphics[width=\textwidth]{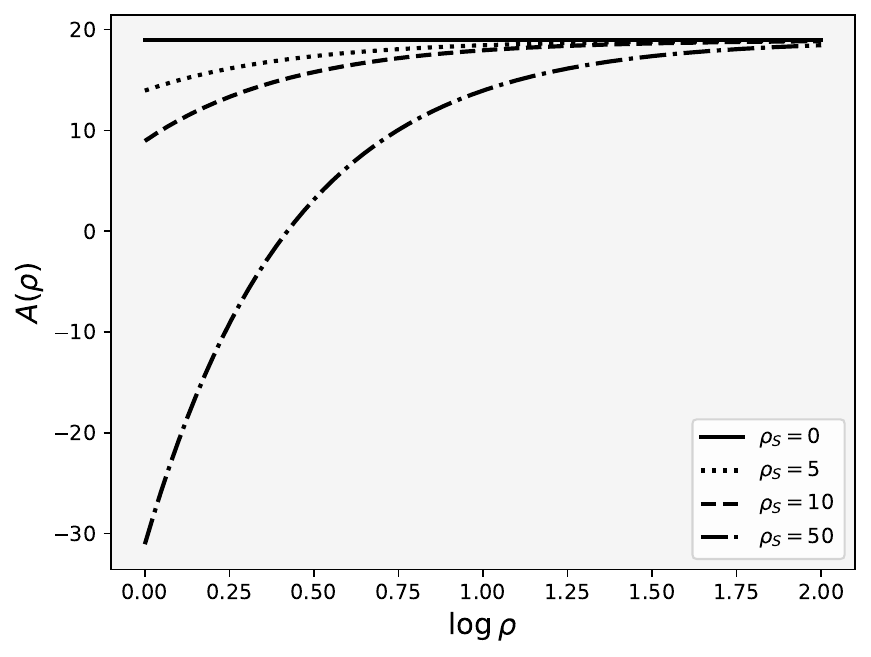}
			\caption{Plot of $A(\rho)$ for varying $\rho_{\scaleto{S}{4pt}}$, with $\alpha_{\scaleto{Q}{4pt}} = 0$, $\overline{a} = 10$, $l = \rho_{\scaleto{obs.}{4pt}} = 3.8 \times 10^{26}$ m, and $N_{\scaleto{Q}{4pt}} = 8.95$.}
			\label{figure-R_s_variando}
		\end{minipage}
	\end{figure}

	From the expression \eqref{Function_A(rho)}, it is clear that the cloud of strings parameter $\overline{a}$ does only vertical shifts, which affect the size of the associated event horizon. We show this behavior through \cref{figure-a_bar_variando}.
	
	Investigating further, we discuss the role of $\rho_{\scaleto{S}{4pt}}$. The Schwarzschild-like radius regulates at which value of $\rho$ the function $A(\rho)$ becomes zero. That is, it is directly proportional to the magnitude of the event horizon ($\rho_{\scaleto{+}{4pt}}$). The bigger is $\rho_{\scaleto{S}{4pt}}$, the larger is $\rho_{\scaleto{+}{4pt}}$. This can be seen in \cref{figure-R_s_variando}.

\begin{figure}[ht!]
		\centering
		\label{Figures}
		\begin{minipage}[t]{0.48\textwidth}
			\includegraphics[width=\textwidth]{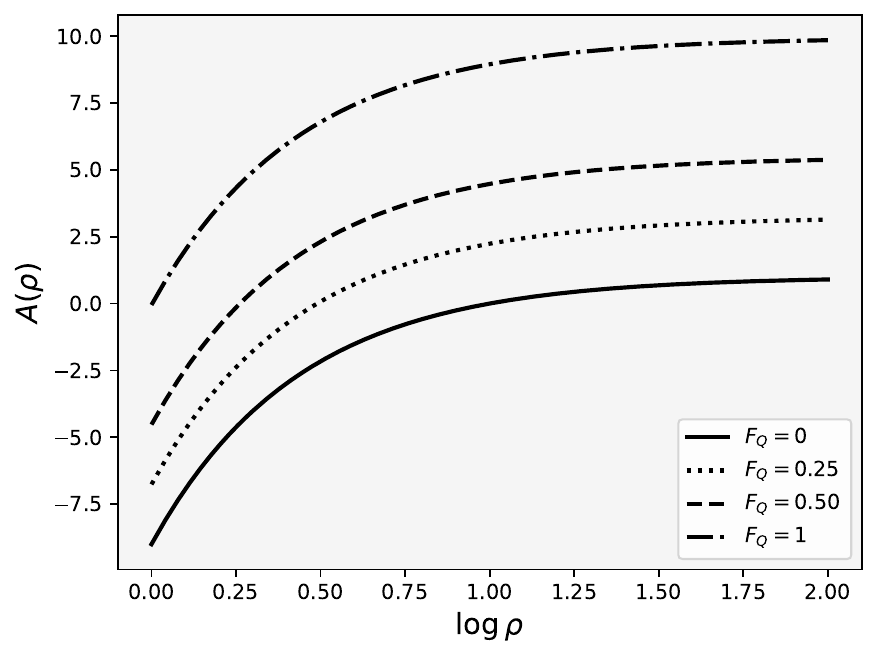}
			\label{fig:Quintessence_Fraction_0}
		\end{minipage}
		\hfill
		\begin{minipage}[t]{0.48\textwidth}
			\includegraphics[width=\textwidth]{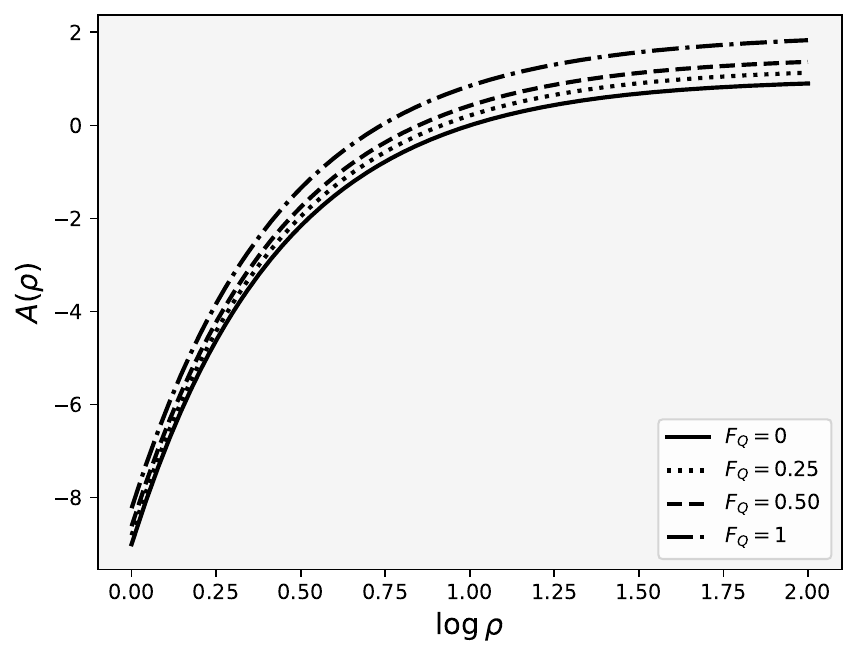}
			\label{fig:Quintessence_Fraction_0-02}
		\end{minipage}
		\caption{$A(\rho)$ for different values of the quintessence fraction $\Omega_{\scaleto{Q}{4pt}}$. The other parameters were set as $\overline{a} = 1$, $\rho_{\scaleto{S}{4pt}} = 10$ m, $l = \rho_{\scaleto{obs.}{4pt}} = 3.8 \times 10^{26}$ m. In the left image, we have the case of $\alpha_{\scaleto{Q}{4pt}} = 0$, while on the right $\alpha_{\scaleto{Q}{4pt}} = 0.02$. Note that any small increase in the value of $\alpha_{\scaleto{Q}{4pt}}$ causes the lines on the graph of $A(\rho)$ to become much closer together, reducing the role played by the quintessence fraction.}
		\label{figure-quintessence-fraction}
	\end{figure}

	Since the contribution from the state parameter $\alpha_{\scaleto{Q}{4pt}}$ and its associated $N_{\scaleto{Q}{4pt}}$ were examined in the previous section, we now proceed to the last factor impacting $A(\rho)$. Previously, we only considered the possibility of $N_{\scaleto{Q}{4pt}}$ taking values up to its highest estimate. However, we could in principle regulate the \enquote{intensity} of this dark energy component for a fixed value of $\alpha_{\scaleto{Q}{4pt}}$ by defining a quintessence fraction $F_{\scaleto{Q}{4pt}}$, given by
	\begin{equation*}
		F_{\scaleto{Q}{4pt}} = \frac{N_{\scaleto{Q}{4pt}}}{N_{\scaleto{Q}{4pt}}^{\text{max}}}\,,
	\end{equation*}
	As illustrated in \cref{figure-quintessence-fraction}, simulations demonstrate that, except for the case $\alpha_{\scaleto{Q}{4pt}} = 0$ (where quintessence contributes most significantly to the metric), values of $N_{\scaleto{Q}{4pt}}$ below those determined by \cref{Estimate-N_Q} are not of interest, especially as they would diminish the overall dark energy content of the universe. Consequently, we will restrict our analysis to $N_{\scaleto{Q}{4pt}}$ values as specified in \cref{Estimate-N_Q}.	
	
	\section{Existence of an event horizon}\label{sec-Event-Horizon}
	
	In \cref{sec-Estimates}, the range and physical interpretation of the parameters considered in this work have been studied. Thinking in terms of these results, we can now explore how they may affect the structure of the spacetime. 
	
	To investigate the formation of an event horizon in a system comprising a black string, quintessence fluid, and a cloud of strings, we seek solutions of $A(\rho_{\scaleto{+}{4pt}})=0$, that is
	\begin{equation}
		A(\rho_{\scaleto{+}{4pt}}) = \overline{a} + \frac{\rho_{\scaleto{+}{4pt}}^{2}}{l^{2}} - \frac{\rho_{\scaleto{S}{4pt}}}{\rho_{\scaleto{+}{4pt}}} + N_{\scaleto{Q}{4pt}}\,\rho_{\scaleto{+}{4pt}}^{2\alpha_{\scaleto{Q}{4pt}}} = 0\,.\label{rho_+}
	\end{equation}
	Since the state parameter $\alpha_{\scaleto{Q}{4pt}}$ can take any value in the interval $0<\alpha_{\scaleto{Q}{4pt}}<1$, the solution to \eqref{rho_+} can vary according to it. 
	Here, we explore specific cases: $\alpha_{\scaleto{Q}{4pt}} = 0,\,1/2,\,1$, with nontrivial values for $\overline{a}$, $\rho_{\scaleto{S}{4pt}}$, and $N_{\scaleto{Q}{4pt}}$. Additionally, we consider the scenarios where quintessence is absent ($N_{\scaleto{Q}{4pt}} = 0$) as well as those where we \enquote{turn off} the clouds of strings (setting $\overline{a} = 0$), which we call \enquote{cloudless}. Finally, as indicated by equation \eqref{rho_+}, the absence of the black string ($\rho_{\scaleto{S}{4pt}} = 0$) results in a system without an event horizon, which we refer to as \enquote{horizonless}.
	
	\subsection{Event horizon in the lower bound}\label{Horizon_Alpha_0}
	
\begin{figure}[ht!]
		\centering
		\begin{minipage}[b]{0.45\textwidth}
			\centering
			\includegraphics[width=\textwidth]{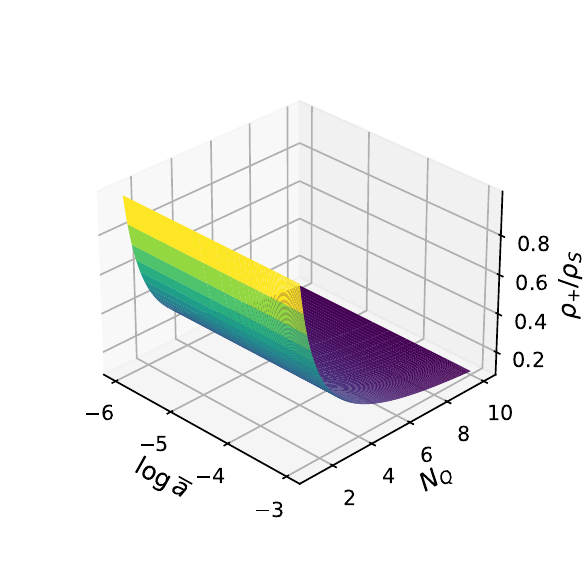}
		\end{minipage}
		\begin{minipage}[b]{0.45\textwidth}
			\centering
			\includegraphics[width=\textwidth]{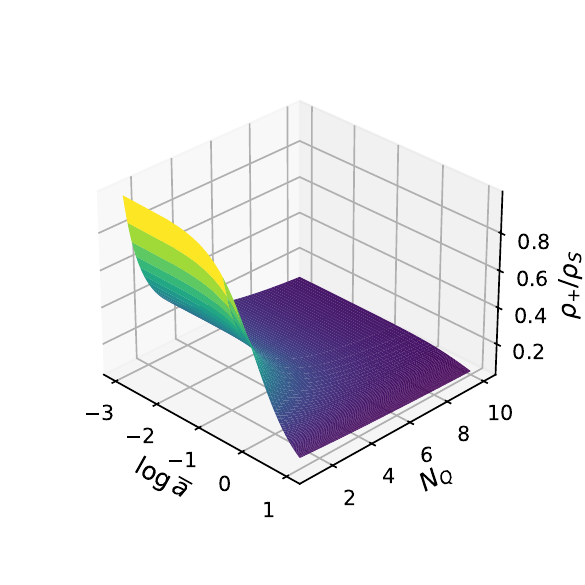}
		\end{minipage}
		\caption{$\rho_{\scaleto{+}{4pt}}/\rho_{\scaleto{S}{4pt}}$ for $\alpha_{\scaleto{Q}{4pt}}=0$ and for $N_{\scaleto{Q}{4pt}} \in (1,\,10)$. On the left, we have $\overline{a}\in (10^{-6}, 10^{-3})$, while on the right $\overline{a} \in (10^{-3}, 10^{1})$.}\label{Rho_Plus_Alpha_0_2}
	\end{figure}

	Let us consider the general solution to the case of $\alpha_{\scaleto{Q}{4pt}}=0$, where \cref{rho_+} implies that
	\begin{equation}
		\left(\overline{a} + N_{\scaleto{Q}{4pt}}\right) + \frac{\rho_{\scaleto{+}{4pt}}^{2}}{l^{2}} - \frac{\rho_{\scaleto{S}{4pt}}}{\rho_{\scaleto{+}{4pt}}} = 0\,.\label{event-alpha-0}
	\end{equation}
	The equation \eqref{event-alpha-0} has a single real root, representing the radius of the event horizon $\rho_{\scaleto{+}{4pt}}$. This root is given by
	\begin{align}
		\rho_{\scaleto{+}{4pt}} &= \xi^{\,1/3} - \frac{l^{2}\left(\overline{a} + N_{\scaleto{Q}{4pt}}\right)}{3}\,\,\xi^{-1/3}\,,\label{Event_Horizon_Alpha_Q_0_Raiz}
	\end{align}
	where
	\begin{equation*}
		\xi = \frac{1}{2} \left[\frac{{{l}^{2}} \sqrt{27 \rho_{\scaleto{S}{4pt}}^{2} + 4\,\left(\overline{a} + N_{\scaleto{Q}{4pt}}\right)^{3}\,l^{2}}}{3^{3/2}} +l^{2}\,\rho_{\scaleto{S}{4pt}}\right]\,.
	\end{equation*}
	However, we can simplify this result under the assumption that we are searching for an event horizon $\rho_{\scaleto{+}{4pt}}$ that is far smaller than the absolute value of the AdS radius $|l|$, that is, $\rho_{\scaleto{+}{4pt}}\ll |l|$. Thus, the contribution from $\rho_{\scaleto{+}{4pt}}^{2}/l^{2} \ll 1$ can be neglected, implying that \cref{event-alpha-0} gives
	\begin{equation}
		\rho_{\scaleto{+}{4pt}} = \frac{\rho_{\scaleto{S}{4pt}}}{\left(\overline{a} + N_{\scaleto{Q}{4pt}}\right)}\,.\label{EH_Alpha_0_Simplified}
	\end{equation}
	For reasonable values of $\overline{a}$ and $N_{\scaleto{Q}{4pt}}$, the event horizon radius $\rho_{\scaleto{+}{4pt}}$ behaves according to \cref{Rho_Plus_Alpha_0_2}.
	It shows that increasing values of $\overline{a}$ and $N_{\scaleto{Q}{4pt}}$ have the role of shrinking the event horizon radius $\rho_{\scaleto{+}{4pt}}$. Conversely, a larger black string contribution (increased $\rho_{\scaleto{S}{4pt}}$) leads to a larger event horizon radius $\rho_{\scaleto{+}{4pt}}$.
	
	\subsection{Event horizon in the middle of the interval}\label{Horizon-Alpha_Meio}
	
\begin{figure}[ht]
		\centering
		\includegraphics[width=0.5\textwidth]{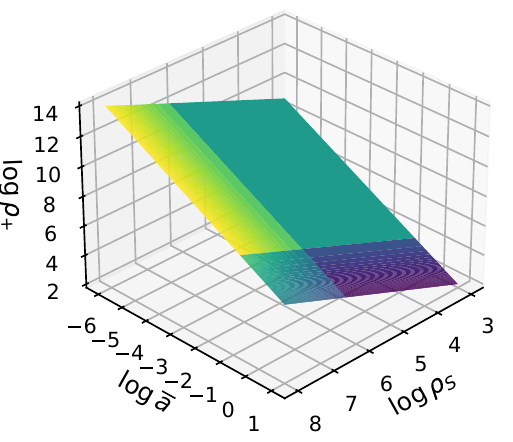}
		\caption{Logarithmic plot of the event horizon radius $\rho_{\scaleto{+}{4pt}}$ as a function of the string cloud parameter $\overline{a}$ and the Schwarzschild-like radius $\rho_{\scaleto{S}{4pt}}$. For illustrative purposes, we consider $10^{-6} \leq \overline{a} \leq 10^{1}$, while $10^{3}\,\text{m} \leq \rho_{\scaleto{S}{4pt}} \leq 10^{8} \,\text{m}$. It is clear from the image and from \cref{Event_Horizon_Simplest} that, the more intense the string clouds, the smaller the event horizon radius.}
		\label{Event_Horizon_Middle}
	\end{figure}

	When $\alpha_{\scaleto{Q}{4pt}}$ lies in the middle of its interval, i.e. $\alpha_{\scaleto{Q}{4pt}} = 1/2$, \cref{rho_+} becomes
	\begin{equation}
		\overline{a} + \frac{\rho_{\scaleto{+}{4pt}}^{2}}{l^{2}} - \frac{\rho_{\scaleto{S}{4pt}}}{\rho_{\scaleto{+}{4pt}}} + N_{\scaleto{Q}{4pt}}\,\rho_{\scaleto{+}{4pt}} = 0\,.\label{Condition_Middle_of_the_Interval}
	\end{equation}
	The solution to the above equation has a single real root, given by
	\begin{equation}
		\rho_{\scaleto{+}{4pt}} = \Xi^{1/3} + \left(\frac{{N_{\scaleto{Q}{4pt}}}^{2}\,l^{4}}{9} - \frac{{l^{2}} \overline{a}}{3}\right)\Xi^{-1/3} - \frac{ N_{\scaleto{Q}{4pt}}\,{l}^{2}}{3}\,,\label{Event_Horizon_Alpha_Q_Meio_Raiz}
	\end{equation}
	where
	\begin{align*}
		\Xi &= \frac{1}{2} \left[\frac{l^{2}\,\sqrt{27 {\rho_{\scaleto{S}{4pt}}}^{2} + \left(18 N_{\scaleto{Q}{4pt}}\,\overline{a}\, l^{2} - 4 {N_{\scaleto{Q}{4pt}}}^{3}\,l^{4}\right)  \rho_{\scaleto{S}{4pt}} - {N_{\scaleto{Q}{4pt}}}^{2}\, \overline{a}^{2}\,l^{4} + 4\, {\overline{a}}^{3}\,l^{2}}}{3^{3/2}} \right.\\
		&\left. + \frac{3 {{l}^{2}} {\rho_{\scaleto{S}{4pt}}} + N_{\scaleto{Q}{4pt}}\,{l}^{4}\,\overline{a}}{3}
		- \frac{2\,{{N_{\scaleto{Q}{4pt}}}}^{3} \,{l}^{6}}{27}\right]\,.
	\end{align*}
	While the full solution to equation \eqref{Condition_Middle_of_the_Interval} is intricate, we can simplify it by considering the case where the event horizon radius $\rho_{\scaleto{+}{4pt}}$ is significantly smaller than the AdS radius $|l|$ , such as $\rho_{\scaleto{+}{4pt}} = 10^{12}$ m, while $|l| = 3.8 \times 10^{26}$ m, which implies $\rho_{\scaleto{+}{4pt}}/|l| \approx 2.6 \times 10^{-14}$. Under this assumption, the term $\rho_{\scaleto{+}{4pt}}^{2}/l^{2}$ becomes negligible, leading to the following simplified expression
	\begin{equation}
		\rho_{\scaleto{+}{4pt}} = \frac{\sqrt{4 {N_{\scaleto{Q}{4pt}}}\, {\rho_{\scaleto{S}{4pt}}}+{{\overline{a}}^{2}}} \pm \overline{a}}{2 {N_Q}}\,.\label{Event_Horizon_Alpha_0.5_Complicated}
	\end{equation}
	With the above result, we can write
	\begin{equation}
		\frac{N_{\scaleto{Q}{4pt}}\,\rho_{\scaleto{+}{4pt}}}{\overline{a}} = \frac{1}{2}\left(\sqrt{1 + \frac{4\,N_{\scaleto{Q}{4pt}}\, \rho_{\scaleto{S}{4pt}}}{\overline{a}^{2}}} - 1\right)\,.\label{N_Q_R+_a_bar}
	\end{equation}
	Then, according to our estimates for $N_{\scaleto{Q}{4pt}}$ when $\alpha_{\scaleto{Q}{4pt}}=1/2$, see \cref{N_Q}, the maximum contribution coming from $N_{\scaleto{Q}{4pt}}$ is of the order of $10^{-26}\,\text{m}^{-1}$, implying it is reasonable to investigate the cases where $N_{\scaleto{Q}{4pt}} \,\rho_{\scaleto{S}{4pt}}/\overline{a}^{2} \ll 1$, 
	\begin{equation}
		\frac{N_{\scaleto{Q}{4pt}} \,\rho_{\scaleto{+}{4pt}}}{\overline{a}} = \frac{N_{\scaleto{Q}{4pt}} \,\rho_{\scaleto{S}{4pt}}}{\overline{a}^{2}} + \mathcal{O}\left(\frac{N_{\scaleto{Q}{4pt}} \rho_{\scaleto{S}{4pt}} }{\overline{a}^{2}}\right)^{2}\,,
	\end{equation}
	and in this regime
	\begin{equation}
		\rho_{\scaleto{+}{4pt}} \approx \frac{\rho_{\scaleto{S}{4pt}}}{\overline{a}}\,.\label{Event_Horizon_Simplest}
	\end{equation}
	From the above result, we see that the intensity of the string clouds, described by $\overline{a}$, is responsible for enlarging or shrinking (when $\overline{a}$ decreases or increases, respectively) the event horizon, as long as our Schwarzschild-like radius remains fixed. Furthermore, the radius of the event horizon $\rho_{\scaleto{+}{4pt}}$ is directly proportional to $\rho_{\scaleto{S}{4pt}}$, the term that quantifies the contribution of the black string. This indicates that as the black string's influence increases (larger $\rho_{\scaleto{S}{4pt}}$), the event horizon expands accordingly. We show the behavior of this scenario in \cref{Event_Horizon_Middle}.	

	\subsection{Event horizon in the upper bound}\label{Event_Horizon_Alpha_Q_1}
	
	In the limiting case where $\alpha_{\scaleto{Q}{4pt}} = 1$, \cref{rho_+} becomes
	\begin{equation}
		N_{\scaleto{QL}{4pt}}\,\rho_{\scaleto{+}{4pt}}^{3}
		+ \overline{a}\,\rho_{\scaleto{+}{4pt}}
		-\rho_{\scaleto{S}{4pt}} = 0\,,\label{Eq_Rho_Plus_Alpha_1}
	\end{equation}
	where the parameter $N_{\scaleto{QL}{4pt}}$ describes the sum of the quintessence parameter ($N_{\scaleto{Q}{4pt}}$) with the cosmological constant ($\Lambda = -3/l^{2}$) according to
	\begin{equation}
		N_{\scaleto{QL}{4pt}} = N_{\scaleto{Q}{4pt}} + \frac{1}{l^{2}}\,.\label{N_QL_Definition}
	\end{equation}
	We can see that, in this regime, the contribution from quintessence to $A(\rho)$ acts in the same form of the cosmological constant, except that $N_{\scaleto{Q}{4pt}}\geq 0$ while $\Lambda$ is negative. As with the previous equations, \cref{Eq_Rho_Plus_Alpha_1} admits only one real root, which is
	\begin{equation}
		\rho_{\scaleto{+}{4pt}} = \Delta^{1/3} - \frac{\overline{a}}{3\, N_{\scaleto{QL}{4pt}}}\,\Delta^{-1/3}\,,\label{Event_Horizon_Alpha_Q_1_Raiz}
	\end{equation}
	with
	\begin{equation*}
		\Delta = \left(\frac{\overline{a}}{3\,N_{\scaleto{QL}{4pt}}}\right)^{3/2}
		\left[\sqrt{1 + \frac{27}{4}\,\frac{N_{\scaleto{QL}{4pt}}\,\rho_{\scaleto{S}{4pt}}^{2}}{\overline{a}^{3}}} + \sqrt{\frac{27}{4}\,\frac{N_{\scaleto{QL}{4pt}}\,\rho_{\scaleto{S}{4pt}}^{2}}{\overline{a}^{3}}}\right]
	\end{equation*}
	In the physical regime where $N_{\scaleto{QL}{4pt}}\,\rho_{\scaleto{S}{4pt}}^{2}/\overline{a}^{3} \ll 1$, we can approximate $\Delta^{1/3}$ as
	\begin{equation}
		\Delta^{1/3} = \left(\frac{\overline{a}}{3\,N_{\scaleto{QL}{4pt}}}\right)^{1/2}
		\left[1 + \frac{1}{3}\, \sqrt{\frac{27}{4}\,\frac{N_{\scaleto{QL}{4pt}}\,\rho_{\scaleto{S}{4pt}}^{2}}{\overline{a}^{3}}} + \frac{3}{8}\,\frac{N_{\scaleto{QL}{4pt}}\,\rho_{\scaleto{S}{4pt}}^{2}}{\overline{a}^{3}}\right]\,, \label{Delta_1-3}
	\end{equation}
	and similarly, we see that
	\begin{equation}
		\Delta^{-1/3} = \left(\frac{\overline{a}}{3\,N_{\scaleto{QL}{4pt}}}\right)^{-1/2}
		\left[1 - \frac{1}{3}\, \sqrt{\frac{27}{4}\,\frac{N_{\scaleto{QL}{4pt}}\,\rho_{\scaleto{S}{4pt}}^{2}}{\overline{a}^{3}}} + \frac{3}{8}\,\frac{N_{\scaleto{QL}{4pt}}\,\rho_{\scaleto{S}{4pt}}^{2}}{\overline{a}^{3}}\right]\,.\label{Delta_-1-3}	
	\end{equation}
	Next, using the results from \cref{Delta_1-3,Delta_-1-3} into \cref{Event_Horizon_Alpha_Q_1_Raiz}, we conclude that
	\begin{equation}
		\rho_{\scaleto{+}{4pt}} = \frac{\rho_{\scaleto{S}{4pt}}}{\overline{a}}\,.\label{EH_Alpha_Q_1}
	\end{equation}
	An alternative reasoning to obtain this result is the following: since we are searching for an event horizon that is much smaller than the observable Universe radius (where the contribution from $N_{\scaleto{QL}{4pt}}$ is non-negligible), we can neglect the term $N_{\scaleto{QL}{4pt}} \rho_{\scaleto{+}{4pt}}^{3}$ in \eqref{Eq_Rho_Plus_Alpha_1} and solve the simplified equation to determine $\rho_{\scaleto{+}{4pt}}$. This approach yields the same result as \cref{EH_Alpha_Q_1}, and is a good approximation for all $\alpha_{\scaleto{Q}{4pt}} > 1/2$, because $N_{\scaleto{Q}{4pt}}$ is infinitely small in this interval.

	\subsection{Additional scenarios}\label{subsection-Additional-scenarios}
	
	Beyond exploring the existence of an event horizon in the regimes where $\alpha_{\scaleto{Q}{4pt}} = 0,\,1/2,\,1$, with $\rho_{\scaleto{S}{4pt}} \neq 0$, $\overline{a} \neq 0$, and $N_{\scaleto{Q}{4pt}} \neq 0$, we investigate the remaining cases where one or more parameters are null. However, note that, as a fundamental requirement for event horizon formation, the black string's associated parameter $\rho_{\scaleto{S}{4pt}}$ must be non-zero. Without a black-string contribution, the system cannot develop an event horizon.
	
	First, we shall explore the cloudless scenarios, where the cloud parameter $\overline{a}$ is set to zero. Then, \cref{rho_+} becomes
	\begin{equation*}
		\frac{\rho_{\scaleto{+}{4pt}}^{2}}{l^{2}} - \frac{\rho_{\scaleto{S}{4pt}}}{\rho_{\scaleto{+}{4pt}}} + N_{\scaleto{Q}{4pt}}\,\rho_{\scaleto{+}{4pt}}^{2\alpha_{\scaleto{Q}{4pt}}} = 0\,.
	\end{equation*}
	If $\alpha_{\scaleto{Q}{4pt}} = 0$, $\rho_{\scaleto{+}{4pt}}$ is just given by \cref{Event_Horizon_Alpha_Q_0_Raiz} setting $\overline{a} = 0$. When $\alpha_{\scaleto{Q}{4pt}} = 1/2$, the result \eqref{Event_Horizon_Alpha_Q_Meio_Raiz} is valid for $\overline{a} = 0$. Nevertheless, if we consider again that $\rho_{\scaleto{+}{4pt}}\ll |l|$, we could simplify these expressions to $\rho_{\scaleto{+}{4pt}} = \rho_{\scaleto{S}{4pt}}/N_{\scaleto{QL}{4pt}}$ (when $\alpha_{\scaleto{Q}{4pt}}=0$) and to	
	\begin{equation}
		\rho_{\scaleto{+}{4pt}} = \left(\frac{\rho_{\scaleto{S}{4pt}}}{N_{\scaleto{Q}{4pt}}}\right)^{1/2}\,,
	\end{equation}
	when $\alpha_{\scaleto{Q}{4pt}} = 1/2$. Our last cloudless case occurs when $\alpha_{\scaleto{Q}{4pt}} = 1$, and its exact form is simply
	\begin{equation}
		\rho_{\scaleto{+}{4pt}} = \left(\frac{\rho_{\scaleto{S}{4pt}}}{N_{\scaleto{QL}{4pt}}}\right)^{1/3}\,,
	\end{equation} 
	with $N_{\scaleto{QL}{4pt}}$ defined by \cref{N_QL_Definition}.

	In addition to the above cloudless scenarios, we also analyze the case when both string clouds and quintessence are absent ($\overline{a} = N_{\scaleto{Q}{4pt}} = 0$), implying that the event horizon is solely driven by the cosmological constant and is given by\begin{equation}
		\rho_{\scaleto{+}{4pt}} = \left(\rho_{\scaleto{S}{4pt}}\,l^{2}\right)^{1/3}\,,
	\end{equation}
	which is large even when compared to the size of the observable universe.
	
	From these results, we see that removing the string clouds significantly increases the event horizon radius $\rho_{\scaleto{+}{4pt}}$  (by several orders of magnitude). Thus, one must pay attention if the approximation for $\rho_{\scaleto{+}{4pt}} \ll |l|$ is still valid.
	
	Our final scenario is the one where quintessence is absent ($N_{\scaleto{Q}{4pt}} = 0$) but the parameters describing the cloud of strings and the black string are non-trivial. The full solution to this case is given by \eqref{Event_Horizon_Alpha_Q_1_Raiz}, setting $N_{\scaleto{Q}{4pt}} = 0$. However, since $\overline{a} \neq 0$ in this case, we find that a good approximation for $\rho_{\scaleto{+}{4pt}}$ is $\rho_{\scaleto{+}{4pt}} = \rho_{\scaleto{S}{4pt}}/\overline{a}$. 
	
	Observing the results presented in this section, it is easy to notice that a wide variety of scenarios is available, and each one of them will lead to a different structure of the spacetime. In the next section, we will study how quantum particles may be inserted in this framework.
	
	\section{The Klein-Gordon equation}\label{sec-KG-equation}
	
	As was pictured in the last section, a system presenting effects of a mass distribution, event horizon, and dark energy may present different kinds of behavior depending on the value of the parameters. So, now it is interesting to investigate how the resulting geometry may affect quantum particles and, in particular, the role of quintessence. As a first example, in this paper, we will consider spin-0 particles.     
	To understand the effects of this spacetime metric, associated with a black string, clouds of strings, and quintessence on the behavior of these particles, we shall write the Klein-Gordon equation using the metric defined by equations \eqref{metric} and \eqref{Function_A(rho)}, and then solve it.
	
	For an arbitrary space-time, associated with a metric tensor $g_{\mu\nu}$, the Klein-Gordon equation is given by
	\begin{equation}
		\frac{1}{\sqrt{-g}}\,\partial_{\mu}\left(g^{\mu\nu}\sqrt{-g}\,\partial_{\nu}\,\phi\right) + \frac{m_{\phi}^{2}\, c^{2}}{\hbar^{2}}\,\phi = 0\,,\label{Klein-Gordon-equation}
	\end{equation}
	where $g=\text{det}(g_{\mu\nu})$ and $m_{\phi}$ is a mass term for the scalar field $\phi$. For the metric in equation \eqref{metric}, the Klein-Gordon equation \eqref{Klein-Gordon-equation} becomes
	\begin{align}
		\frac{1}{A(\rho)}\,\partial_{t}^{2}\,\phi - \frac{1}{\rho^{2}}\,\partial_{\rho}\left[\rho^{2}A(\rho)\,\partial_{\rho}\,\phi\right] - \frac{1}{\rho^{2}}\left[\partial_{\varphi}^{2} - l^{2}\partial_{z}^{2}\right]\phi + \overline{m}_{\phi}^{2}\,\phi = 0\,,\label{complete-PDE}
	\end{align}
	where $\overline{m}_{\phi}^{2} = m_{\phi}^{2} c^{2}/\hbar^{2}$. 
	
	To separate the variables, we propose that $\phi = T(t)R(\rho)\Phi(\varphi)Z(z)$, and
	plugging this ansatz into \eqref{complete-PDE} yields
	\begin{equation}
		\phi = \exp\left({-i\epsilon t}\right) \exp\left({in\varphi}\right) \exp\left({izp_{z}/l}\right)
		\,R(\rho)\,,\label{ansatz-campo-escalar}
	\end{equation} 
	where $n \in \mathbb{Z}$ and $\epsilon,\,p_{z}\in\mathbb{R}$. The ordinary differential equation for $R(\rho)$ is
	\begin{equation}
		\partial_{\rho}^{2}\,R(\rho) + \chi(\rho)\,\partial_{\rho}\,R(\rho) + \tau(\rho)\,R(\rho) = 0\,,\label{general-equation-for-R}
	\end{equation}
	with
	\begin{align}
		\chi(\rho) &\coloneqq \frac{1}{A(\rho)}\,\partial_{\rho}A + \frac{2}{\rho}\,,\label{chi}\\
		\tau(\rho) &\coloneqq \frac{1}{A(\rho)}\left(\frac{\epsilon^{2}}{A(\rho)} - \frac{n^{2} + p_{z}^{2}}{\rho^{2}} - \frac{m_{\phi}^{2}\,c^{2}}{\hbar^{2}}\right)\,,\label{tau}
	\end{align}
	where $A(\rho)$ is the solution given by \cref{Function_A(rho)}.
	
	To solve for $R(\rho)$ we use the strategy of writing \eqref{general-equation-for-R} in the Liouville normal form \cite{Titchmarsh1962}, introducing the auxiliary function $u(\rho)$ related to $R(\rho)$ by
	\begin{equation}
		R(\rho) = \frac{R_{0}}{\rho \sqrt{A(\rho)}}\,u(\rho)\,,\label{radial-from-u}
	\end{equation}
	where $R_{0}$ is a normalization constant. The resulting normal equation for $u(\rho)$ is
	\begin{equation}
		u''(\rho) + V_{\text{eff}}(\rho)\,u(\rho) = 0\,,\label{eq-geral-u}
	\end{equation}
	where the primes ($"$) stand for derivatives in $\rho$. The effective potential $V_{\text{eff}}(\rho)$ is defined as
	\begin{equation}
		V_{\text{eff}}(\rho) = \frac{A'(\rho)^{2} + 4\epsilon^{2}}{4A(\rho)^{2}} - \frac{1}{A(\rho)}\left[\frac{1}{2}A''(\rho) + \frac{A'(\rho)}{\rho} + \frac{\kappa^{2}}{\rho^{2}} + \overline{m}_{\phi}^{2}\right]\,,\label{Effective-Potential}
	\end{equation}
	where $\kappa^{2} \coloneqq p_{z}^{2} + n^{2}$.
	
	The solutions to $R(\rho)$ depend on the scenario under consideration for this system, the existence of an event horizon, mass distribution, and other possibilities discussed in the last section.

	\subsection{Implementing a new coordinate system}
	
	As a first possibility, let us assume a set of parameters that determines the existence of an event horizon $\rho_{\scaleto{+}{4pt}}\neq 0$. Then, we define a new dimensionless coordinate $x$, according to
	\begin{equation}
		x = \frac{\rho}{\rho_{\scaleto{+}{4pt}}}\,,\label{new-coordinate}
	\end{equation}
	with $x \in \left(0,\,+\infty\right)$. We search for solutions where $x > 1$, i.e., outside the event horizon, to describe observable particles. Those cases where the black string possesses a naked singularity, meaning that $\rho_{\scaleto{S}{4pt}} = 0$, will be analyzed in our next paper.
	
	In terms of $x$, any $n$-th derivative in $\rho$ becomes ${d}^{n}/d{\rho}^{n} = \left(\rho_{\scaleto{+}{4pt}}\right)^{-n}\,{d}^{n}/d {x}^{n}$, so that we can rewrite \cref{eq-geral-u} in terms of $x$. The procedure results in
	\begin{equation}
		u''(x) + \rho_{\scaleto{+}{4pt}}^{2}\,V_{\text{eff}}(x)\,u(x) = 0\,,\label{Eq-para-u}
	\end{equation}
	with
	\begin{equation}
		\rho_{\scaleto{+}{4pt}}^{2}\,V_{\text{eff}}(x) = \frac{A'(x)^{2} + 4\left(\rho_{\scaleto{+}{4pt}}\epsilon\right)^{2}}{4\,A(x)^{2}} 
		- \frac{1}{A(x)}\left[\frac{1}{2}A''(x) + \frac{A'(x)}{x} + \frac{\kappa^{2}}{x^{2}} + \rho_{\scaleto{+}{4pt}}^{2}\,\overline{m}_{\phi}^{2}\right]\,.\label{Potencial_em_x}
	\end{equation}
	It is worth stressing that the derivatives in \cref{Eq-para-u,Potencial_em_x} are with respect to $x$, i.e., $d/d x$. These equations are now in a suitable form for investigating the solutions.

	\subsection{Solutions near the event horizon} \label{Basic_Solution_Near}
	
	Let us solve \cref{Eq-para-u,Potencial_em_x} for $\rho$ in a small vicinity of $\rho_{\scaleto{+}{4pt}}$, which means particles near the event horizon. In this case, it is reasonable to approximate $A(\rho)$, given by \eqref{Function_A(rho)}, according to
	\begin{equation*}
		A(\rho) \approx \left.\frac{dA}{d \rho}\right|_{\rho_{\scaleto{+}{4pt}}}\left(\rho - \rho_{\scaleto{+}{4pt}}\right)\,,
	\end{equation*}
	and under the definition
	\begin{equation*}
		\beta_{\scaleto{+}{4pt}} \coloneqq \rho_{\scaleto{+}{4pt}}\left.\frac{dA}{d \rho}\right|_{\rho_{\scaleto{+}{4pt}}}\,,\label{Beta+}
	\end{equation*}
	it follows directly from \eqref{Function_A(rho)} that
	\begin{equation}
		\beta_{\scaleto{+}{4pt}} = \frac{2 \rho_{\scaleto{+}{4pt}}^{2}}{l^{2}} + \frac{\rho_{\scaleto{S}{4pt}}}{\rho_{\scaleto{+}{4pt}}} + 2\alpha_{\scaleto{Q}{4pt}}\,N_{\scaleto{Q}{4pt}}\,\rho_{\scaleto{+}{4pt}}^{2\alpha_{\scaleto{Q}{4pt}}}\,,\label{Beta_Plus}	
	\end{equation}
	so that $\beta_{\scaleto{+}{4pt}}>0$, and we require $\rho_{\scaleto{S}{4pt}} \neq 0$ to promote $\rho_{\scaleto{+}{4pt}} \neq 0$. 
	Thus, the function $A(\rho)$ can be approximated by $A(\rho) \approx \beta_{\scaleto{+}{4pt}}\left(\rho/\rho_{\scaleto{+}{4pt}} - 1\right)$, which can be directly converted to
	\begin{equation}
		A(x) = \beta_{\scaleto{+}{4pt}}\left(x - 1\right)\,.\label{A(x)-type-1-solution}
	\end{equation}
	Next, we plug $A(x)$ together with its derivatives $A'(x) = \beta_{\scaleto{+}{4pt}}$ and $A''(x) = 0$ into \cref{Potencial_em_x} to show that
	\begin{equation}
		\rho_{\scaleto{+}{4pt}}^{2}\,V_{\text{eff}}(x) = \frac{1+\kappa^{2}/\beta_{\scaleto{+}{4pt}}}{x}
		- \frac{1 + (\kappa^{2} + \rho_{\scaleto{+}{4pt}}^{2}\,\overline{m}_{\phi}^{2})/\beta_{\scaleto{+}{4pt}}}{x-1}
		+ \frac{\kappa^{2}/\beta_{\scaleto{+}{4pt}}}{x^{2}}
		+ \frac{1/4 + \left(\epsilon\rho_{\scaleto{+}{4pt}}/\beta_{\scaleto{+}{4pt}}\right)^{2}}{\left(x-1\right)^{2}}\,,
	\end{equation}
	where we proceeded with a partial fraction decomposition since we intend to give the solution to \cref{Eq-para-u} in terms of the confluent Heun equation \cite{Ronveaux1995,Hortacsu2018}.
	
\begin{figure}[ht]
		\begin{center}
			\begin{minipage}[c]{0.48\textwidth}
				\includegraphics[width=\textwidth]{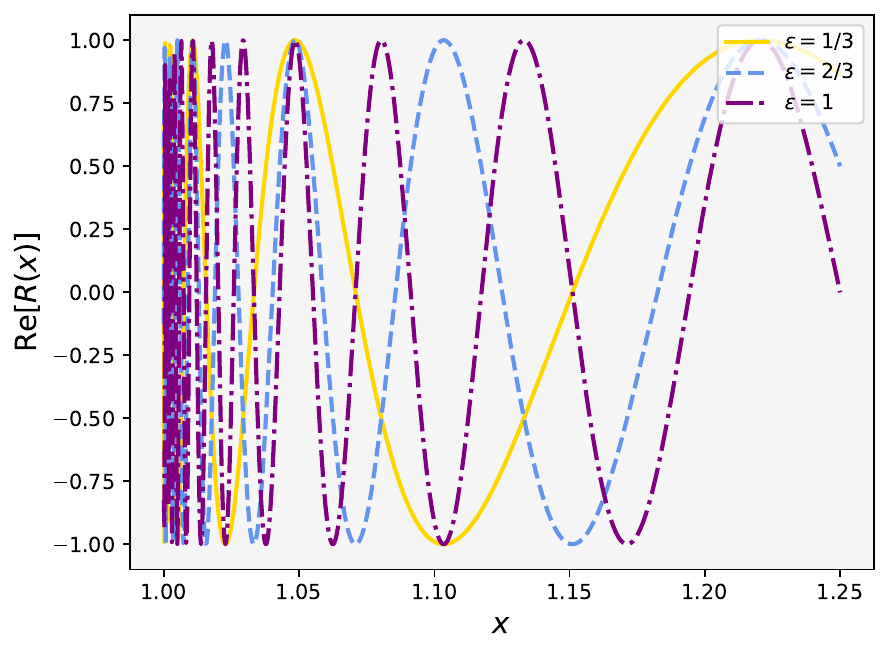}
			\end{minipage}
			\begin{minipage}[c]{0.48\textwidth}
				\includegraphics[width=\textwidth]{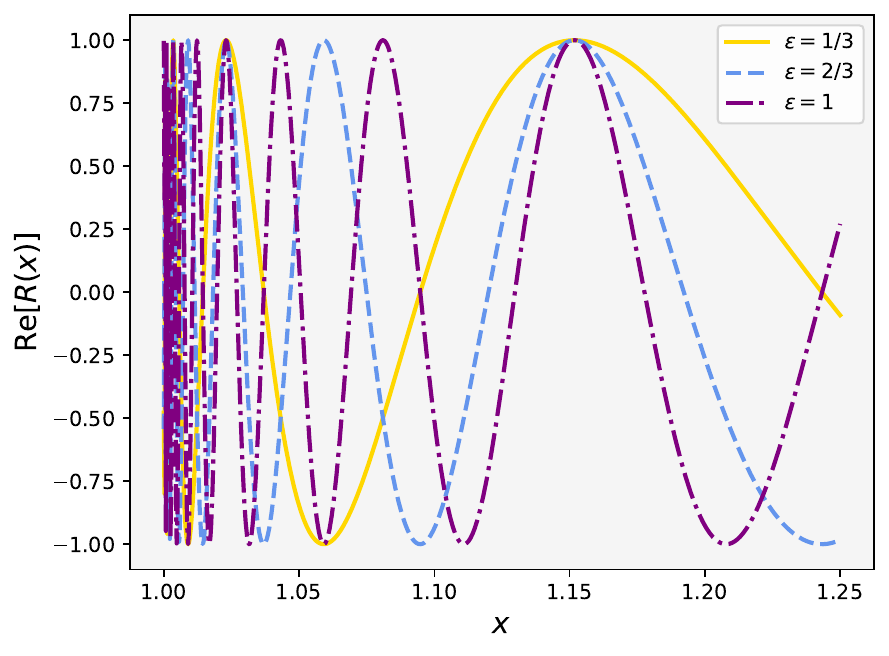}
			\end{minipage}
		\end{center}
		\caption{Plot of the real part of $(x-1)^{\gamma/2}$ for $\alpha_{\scaleto{Q}{4pt}} = 0$ (left) and $\alpha_{\scaleto{Q}{4pt}}=1/2$ (right). In the left image, where $\alpha_{\scaleto{Q}{4pt}}=0$, we considered $\rho_{\scaleto{S}{4pt}} = 10^{3}$ m, $\overline{a} = 10^{-2}$, $N_{\scaleto{Q}{4pt}} = 8.95$. In the right, when $\alpha_{\scaleto{Q}{4pt}} = 1/2$, we used $\rho_{\scaleto{S}{4pt}} = 10^{3}$ m, $\overline{a} = 10^{1}$, and $N_{\scaleto{Q}{4pt}} = 2.35 \times 10^{-26}$ m$^{-1}$. The values of $N_{\scaleto{Q}{4pt}}$ follow from \cref{N_Q} considering $\rho_{\scaleto{obs.}{4pt}} = |l| = 3.8 \times 10^{26}$ m. We have not included the plots for the imaginary components because $\operatorname{Im}[R(x)] = \sin\left[\frac{\epsilon \, \rho_{\scaleto{+}{4pt}}}{\beta_{\scaleto{+}{4pt}}}\ln(x-1)\right]$ only differ by a phase shift of $\pi/2$ from the real case $\operatorname{Re}[R(x)]$.}
		\label{Radial_WF}
	\end{figure}

	Under the comparison with \cref{Eq_u_Normal_Form,Parameters_HeunC_Normal}, we conclude that the solution for $u(x)$ is the following:
	\begin{align*}
		u(x) &= x^{(1+\beta)/2}\,\left(x-1\right)^{(1+\gamma)/2}\,\exp\left(\alpha x/2\right)\, \left[ c_{1}
		\text{HeunC}\left(\alpha,\,\beta,\,\gamma,\,\delta,\,\eta;\,x\right) \right.\\
		&\left.+ c_{2} \, x^{-\beta}\,
		\text{HeunC}\left(\alpha,\,-\beta,\,\gamma,\,\delta,\,\eta;\,x\right)\right]\,, 
	\end{align*}
	while the Heun parameters are
	\begin{equation}
		\begin{aligned}
			\alpha &= 0\,,\\
			\beta &= \pm \sqrt{1 - \frac{4\,\kappa^{2}}{\beta_{\scaleto{+}{4pt}}}}\,,\\
			\gamma &= \pm i\,\frac{2 \epsilon \rho_{\scaleto{+}{4pt}}}{\beta_{\scaleto{+}{4pt}}}\,,\\
			\delta &= -\frac{\rho_{\scaleto{+}{4pt}}^{2}\,\overline{m}_{\phi}^{2}}{\beta_{\scaleto{+}{4pt}}}\,,\\
			\eta &= -\frac{1}{2} - \frac{\kappa^{2}}{\beta_{\scaleto{+}{4pt}}}\,.
		\end{aligned} \label{Heun_Param_1}
	\end{equation}
	As a consequence of \cref{radial-from-u,A(x)-type-1-solution}, the radial part of the wave function for the scalar particle becomes
	\begin{equation}
		\begin{aligned}
			R(x) &= \left(\rho_{\scaleto{+}{4pt}}^{2} \, \beta_{\scaleto{+}{4pt}}\right)^{-1/2}\,x^{(\beta-1)/2}\,\left(x-1\right)^{\gamma/2}\,\left[ c_{1}\,
			\text{HeunC}\left(\alpha,\,\beta,\,\gamma,\,\delta,\,\eta;\,x\right) \right.\\
			&\left.+ c_{2}\, x^{-\beta}\,
			\text{HeunC}\left(\alpha,\,-\beta,\,\gamma,\,\delta,\,\eta;\,x\right)\right]\,,
		\end{aligned} \label{type-1-full-solution}
	\end{equation}
	where $\beta_{\scaleto{+}{4pt}}$, $\rho_{\scaleto{+}{4pt}}$, and $x \in \mathbb{R}$. 	We must remember that the above solution is valid for a small region where $(x-1) \approx 0$, with $x>1$. In this region, the local confluent Heun function can be written according to \cref{HeunCl_around_1}:
	\begin{equation}
		\text{HeunC}\ell\left(\alpha,\,\beta,\,\gamma,\,\delta,\,\eta;\,z-1\right) = \sum_{m = 0}^{+\infty} c_{m} \left(x-1\right)^{m}\,,
		\label{R_xheun}
	\end{equation}
	where $c_{0} = 1$ and the remaining coefficients are determined by \cref{c_1_Confluente_1,Termo_Geral,M_0,M_1,M_2}. Thus, when $x\to 1$, we consider	$\text{HeunC}\left(\alpha,\,\pm\beta,\,\gamma,\,\delta,\,\eta;\,x\right) \approx 1$. In fact, by analyzing the limit $\lim_{x\to 1} R(x)$, we obtain that $R(x \approx 1)$ behaves according to
	\begin{equation}
		R(x) = \left(x-1\right)^{\gamma/2}\,\left[ c_{1}\, x^{\beta/2}
		+ c_{2}\, x^{-\beta/2}\right]\,,
	\end{equation}
	where $c_{1}$ and $c_{2}$ are constants. Note that we have considered $x^{-1/2} \approx 1$ in this domain. From the above result, this radial wave function can be written as
	\begin{equation}
		\begin{aligned}
			R(x) 
			&= c_{1}\,\exp\left[\frac{\gamma}{2} \ln (x-1) + \frac{\beta}{2} (x-1)\right]\\ 
			&+ c_{2}\,\exp\left[\frac{\gamma}{2} \ln (x-1) - \frac{\beta}{2} (x-1)\right]\,.
		\end{aligned}\label{R_Near_x_1}
	\end{equation}
	It is interesting to see that, unless $|\beta| \gg |\gamma|$, the behavior of the wave function is dominated by
	\begin{equation}
		R(x) \sim  \left(x-1\right)^{\gamma/2}\,, \label{R_of_x_Vicinity_1}
	\end{equation}
	up to a multiplicative constant. Thus, considering the expression for $\gamma$ shown in \eqref{Heun_Param_1}, the real part of $R(x)$ would behave as $\cos\left[\frac{\epsilon \, \rho_{\scaleto{+}{4pt}}}{\beta_{\scaleto{+}{4pt}}}\ln(x-1) + \theta_{0}\right]$, where $\theta_{0}\in \mathbb{R}$ is a constant phase. This suggests that the radial wave function oscillates rapidly near the event horizon, as we show in \cref{Radial_WF}. However, we stress that the solution, given by eq. (\ref{type-1-full-solution}), can be investigated in a better way by observing the numerical values of its parameters for each case. This will be done in the next section.
	
	\subsection{Physical scenarios:}
	
	As we can see, the factors $\gamma$ and $\beta$ are fundamental in determining the behavior of the solution shown in eq. (\ref{R_Near_x_1}), in the regime considered in this section. 
	To analyze their implications in equation \eqref{R_Near_x_1}, we explore the physically plausible scenarios regarding $\overline{a}$, $\rho_{\scaleto{S}{4pt}}$, and $N_{\scaleto{Q}{4pt}}$ and their constraints. This analysis will reveal how $\rho_{\scaleto{+}{4pt}}$ and $\beta_{\scaleto{+}{4pt}}$ are interconnected. All scenarios have $\rho_{\scaleto{S}{4pt}}\neq 0$, otherwise there would be no event horizon formation. In the standard scenario, when $\overline{a} \neq 0$ and $N_{\scaleto{Q}{4pt}} \neq 0$, we use the results of \cref{sec-Event-Horizon} and of \cref{Beta_Plus} to obtain the set of equations displayed in \cref{table-standard-cases}.
	\begin{table}[ht]
		\begin{center}
			\begin{tabular}{ | c | c | c | c |}
				\hline
				$\alpha_{\scaleto{Q}{4pt}}$ & $\rho_{\scaleto{+}{4pt}}$ & $\beta_{\scaleto{+}{4pt}}$ & $\rho_{\scaleto{+}{4pt}}/\beta_{\scaleto{+}{4pt}}$\\
				\hline\hline
				$0$ & $\rho_{\scaleto{S}{4pt}}/(\overline{a} + N_{\scaleto{Q}{4pt}})$ & $\overline{a} + N_{\scaleto{Q}{4pt}}$ & $\rho_{\scaleto{S}{4pt}}/(\overline{a} + N_{\scaleto{Q}{4pt}})^{2}$\\
				
				$1/2$ & $\rho_{\scaleto{S}{4pt}}/\overline{a}$ & $\overline{a} \left(1 + N_{\scaleto{Q}{4pt}} \rho_{\scaleto{S}{4pt}}/ \overline{a}^{2}\right)$ & $\left(\rho_{\scaleto{S}{4pt}}/\overline{a}^{2}\right)\left(1 - N_{\scaleto{Q}{4pt}} \rho_{\scaleto{S}{4pt}}/\overline{a}^{2}\right)$\\
				
				$1$ & $\rho_{\scaleto{S}{4pt}}/\overline{a}$ & $\overline{a} \left(1 + 2\, N_{\scaleto{QL}{4pt}} \rho_{\scaleto{S}{4pt}}^{2}/ \overline{a}^{3}\right)$ & $\left(\rho_{\scaleto{S}{4pt}}/\overline{a}^{2}\right)\left(1 - 2\,N_{\scaleto{QL}{4pt}} \rho_{\scaleto{S}{4pt}}^{2}/\overline{a}^{3}\right)$\\
				\hline
			\end{tabular}
		\end{center}
		\caption{Results of $\rho_{\scaleto{+}{4pt}}$, $\beta_{\scaleto{+}{4pt}}$, and $\rho_{\scaleto{+}{4pt}}/\beta_{\scaleto{+}{4pt}}$ for various scenarios under the constraints $\rho_{\scaleto{+}{4pt}} \ll |l|$ (constraint for the first and second rows), $N_{\scaleto{Q}{4pt}} \rho_{\scaleto{S}{4pt}}/\overline{a}^{2} \ll 1$ (when $\alpha_{\scaleto{Q}{4pt}} = 1/2$), and $N_{\scaleto{QL}{4pt}} \rho_{\scaleto{S}{4pt}}^{2}/\overline{a}^{3} \ll 1$ (when $\alpha_{\scaleto{Q}{4pt}} = 1$). Note that $\rho_{\scaleto{S}{4pt}}$, $\overline{a}$, and $N_{\scaleto{Q}{4pt}}$ are all non-trivial and that $N_{\scaleto{QL}{4pt}}$ was defined by \cref{N_QL_Definition}.}
		\label{table-standard-cases}
	\end{table}
	Conversely, if $\overline{a}=0$ so that we have a cloudless scenario, we use the results of \cref{subsection-Additional-scenarios} to show that $\rho_{\scaleto{+}{4pt}}$, $\beta_{\scaleto{+}{4pt}}$, and $\rho_{\scaleto{+}{4pt}}/\beta_{\scaleto{+}{4pt}}$ are given by \cref{table-cloudless-cases}.
	\begin{table}[ht]
		\begin{center}
			\begin{tabular}{ | c | c | c | c |}
				\hline
				$\alpha_{\scaleto{Q}{4pt}}$ & $\rho_{\scaleto{+}{4pt}}$ & $\beta_{\scaleto{+}{4pt}}$ & $\rho_{\scaleto{+}{4pt}}/\beta_{\scaleto{+}{4pt}}$\\
				\hline\hline
				$0$ & $\rho_{\scaleto{S}{4pt}}/N_{\scaleto{Q}{4pt}}$ & $N_{\scaleto{Q}{4pt}}$ & $\rho_{\scaleto{S}{4pt}}/N_{\scaleto{Q}{4pt}}^{2}$\\ 
				$1/2$ & $\left(\rho_{\scaleto{S}{4pt}}/N_{\scaleto{Q}{4pt}}\right)^{1/2}$ & $2\left(\rho_{\scaleto{S}{4pt}} N_{\scaleto{Q}{4pt}}\right)^{1/2}$ & $1/(2 N_{\scaleto{Q}{4pt}})$\\
				$1$ & $\left(\rho_{\scaleto{S}{4pt}}/N_{\scaleto{QL}{4pt}}\right)^{1/3}$ & $3 \left(\rho_{\scaleto{S}{4pt}}^{2} N_{\scaleto{QL}{4pt}}\right)^{1/3}$ & $(1/3)\left(\rho_{\scaleto{S}{4pt}} N_{\scaleto{QL}{4pt}}^{2}\right)^{-1/3}$\\
				\hline
			\end{tabular}
		\end{center}
		\caption{Expressions for $\rho_{\scaleto{+}{4pt}}$, $\beta_{\scaleto{+}{4pt}}$, and $\rho_{\scaleto{+}{4pt}}/\beta_{\scaleto{+}{4pt}}$ for cloudless scenarios ($\overline{a} = 0$). In the first two rows, the results follow from the constraint $\rho_{\scaleto{+}{4pt}} \ll |l|$. The third row is an exact solution, where $N_{\scaleto{QL}{4pt}}$ is the sum of $N_{\scaleto{Q}{4pt}}$ with $1/l^{2}$, as defined by \cref{N_QL_Definition}.}
		\label{table-cloudless-cases}
	\end{table}
	Our final scenarios are the absence of quintessence (with $\overline{a}\neq 0$, $\rho_{\scaleto{S}{4pt}}\neq 0$) together with the case where we remove the quintessential fluid and the clouds of strings, i.e., $\overline{a} = 0$ and $N_{\scaleto{Q}{4pt}} = 0$. These are described in \cref{table-supplementary-cases}. However, we should note that these last scenarios are special cases of \cref{table-standard-cases} (first row) and of \cref{table-cloudless-cases} (third row), just setting $N_{\scaleto{Q}{4pt}} = 0$.
	\begin{table}[ht]
		\begin{center}
			\begin{tabular}{| c | c | c |}
				\hline
				$\rho_{\scaleto{+}{4pt}}$ & $\beta_{\scaleto{+}{4pt}}$ & $\rho_{\scaleto{+}{4pt}}/\beta_{\scaleto{+}{4pt}}$\\
				\hline\hline
				$\rho_{\scaleto{S}{4pt}}/\overline{a}$ & $\overline{a}$ & $\rho_{\scaleto{S}{4pt}}/\overline{a}^{2}$\\
				$\left(\rho_{\scaleto{S}{4pt}}\, l^{2}\right)^{1/3}$ & $3 \left(\rho_{\scaleto{S}{4pt}}/l\right)^{2/3}$ & $(1/3) \left(l^{4}/\rho_{\scaleto{S}{4pt}}\right)^{1/3}$\\
				\hline
			\end{tabular}
		\end{center}
		\caption{Equations describing $\rho_{\scaleto{+}{4pt}}$, $\beta_{\scaleto{+}{4pt}}$, and $\rho_{\scaleto{+}{4pt}}/\beta_{\scaleto{+}{4pt}}$. The first scenario is the one without quintessence (first row), under the constraint $\rho_{\scaleto{+}{4pt}}\ll |l|$. The second scenario has the black-string only (second row), with no further constraints.}
		\label{table-supplementary-cases}
	\end{table}
	
	If, for example, we take the value $\alpha_{\scaleto{Q}{4pt}} = 1/2$ and consider the explicit Heun parameters, given by \cref{Heun_Param_1}, with the expressions determined at \cref{table-standard-cases} (second row), we can write the solution \eqref{R_Near_x_1}, with $(x-1) \to 0^{+}$, as
	\begin{equation}
		R(x) = \Phi_{+}\,\exp\left(i\,\delta_{D+}\right) + \Phi_{-}\,\exp\left(i\,\delta_{D-}\right)\,,
	\end{equation}
	where	
	\begin{equation}
		\Phi_{\pm}(x) = c_{\pm}\,\left[\,i \epsilon\, \frac{\rho_{\scaleto{S}{4pt}}}{\overline{a}^{2}}\, \ln(x-1) \pm \sqrt{\frac{1}{4} - \frac{\kappa^{2}}{\overline{a}}}\,(x-1)\right]\, 
	\end{equation}
	are functions of $x$ only (meaning that they are independent from $N_{\scaleto{Q}{4pt}}$), $c_{\pm}$ are constants, and
	\begin{equation}
		\delta_{D \pm} = -\left[\frac{\epsilon \rho_{\scaleto{S}{4pt}}}{\overline{a}}\, \ln (x-1) \pm \frac{i\,\kappa^{2}}{\sqrt{\overline{a}}\sqrt{\overline{a} - 4\kappa^{2}}}\, (x-1)\right] \frac{N_{\scaleto{Q}{4pt}}\,\rho_{\scaleto{S}{4pt}}}{\overline{a}^{2}}\,.
	\end{equation}
	are functions of $x$ where the dependence on the quintessence parameter $N_{\scaleto{Q}{4pt}}$ appears in a factorized form. As long as all the dependence on $N_{\scaleto{Q}{4pt}}$ is contained in these functions, we shall name them the dark phase. Consequently, the study of $\delta_{D\pm}$ provides all the information about the dark energy in the system under consideration and, for this reason, may be used to determine observables. Similar behavior is expected to occur in the other scenarios presented in Tables \ref{table-standard-cases}-\ref{table-supplementary-cases} and also when solving the equation for other regions of $\rho$. This study will be presented in our next paper.

	\section{Conclusion} \label{sec-Conclusion}
	
	In this work, we investigated the impact of quintessence, a potential dark energy candidate, on the spacetime geometry surrounding a black string and a cloud of strings. With this procedure, we modeled a system with mass distribution, event horizon, and dark energy in a way that these contributions could be compared. By solving Einstein's field equations, we obtained a metric that reveals how quintessence, characterized by the parameter $\alpha_{\scaleto{Q}{4pt}}$, modifies the spacetime. To obtain a deeper physical understanding of this system, we estimated admissible values (in SI units) for the parameters $\overline{a}$, $\rho_{\scaleto{S}{4pt}}$, and $N_{\scaleto{Q}{4pt}}$ present in the metric. We have also explored how quintessence would make its greatest contribution to the metric in the limit $\alpha_{\scaleto{Q}{4pt}} \to 0$. Conversely, as $\alpha_{\scaleto{Q}{4pt}}$ increases, the contribution from quintessence to the metric becomes non-negligible only at extremely large distances (of the order of $\rho_{\scaleto{obs.}{4pt}}$).
	
	We examined the existence of an event horizon $\rho_{\scaleto{+}{4pt}}$ for all scenarios when $\alpha_{\scaleto{Q}{4pt}} = 0,\,1/2,$ and $1$. It became evident that the cloud of strings parameter $\overline{a}$ had the role of shrinking the event horizon when compared to \enquote{cloudless} cases. On the other hand, we could see that the presence of quintessence (when $\alpha_{\scaleto{Q}{4pt}} \geq 1/2$) had a subtle influence on $\rho_{\scaleto{+}{4pt}}$. The dominant contribution to $\rho_{\scaleto{+}{4pt}}$ came from the Schwarzschild-like radius and the string clouds parameter.
	
	When solving the Klein-Gordon equation near the event horizon, we have obtained that the radial wave function $R(x)$ of the spin$-0$ particle, where $x = \rho/\rho_+$, behaves as $(x-1)^{\gamma/2}$, where $\gamma = i\,\epsilon \rho_+/\beta_+$. We investigated general results for this regime, including all known cases of event horizon formation for this metric. In special, we defined the concept of \enquote{dark phase} to express the change caused by quintessence in the wave function, and it may be used to build observable objects. The solution of the equations for different sets of parameters and different regions of $\rho$ will be presented in our next work, and with these results it will be possible to investigate the dark phase more extensively.

	Our results provide valuable insights into the interplay between gravity, matter, and dark energy in a cylindrically symmetric universe. Future research could extend these findings to more complex scenarios, such as rotating black strings or alternative quintessence models, to further elucidate the nature of dark energy and its cosmological implications.

	\section*{Acknowledgments}
	
	M.L.D and L.G.B would like to thank the Coordenação de Aperfeiçoamento de Pessoal de Nível Superior – Brasil (CAPES) – Finance Code 001. C.C.B.Jr. thanks the Conselho Nacional de Desenvolvimento Científico e Tecnológico (CNPq) for financial support.

	\appendix
	
	\section{Units of the parameters}\label{ap-units}
	
	To estimate the possible values for $a,\,\rho_{\scaleto{S}{4pt}},\,N_{\scaleto{Q}{4pt}}$, and $m_{\phi}$, as well as to determine their influence in $A(\rho)$, it is interesting to proceed with a dimensional analysis in terms of SI units. 
	
	Let us start with the parameter $a$, from the clouds of strings. Using that $A(\rho)$ must be dimensionless, we see that
	\begin{equation*}
		\left[a\right]_{\scaleto{SI}{4pt}} = \left[\frac{8\pi G}{c^{4}}\right]^{-1}_{\scaleto{SI}{4pt}} = \text{kg} \cdot \text{m}\cdot \text{s}^{-2}\,,
	\end{equation*}
	The dimension of $a$ implies that
	\begin{equation*}
		\left[T_{t}^{\,\,t}\right]_{\scaleto{SI}{4pt}} = \left[\frac{a}{\rho^{2}}\right]_{\scaleto{SI}{4pt}} = \text{kg}\cdot \text{m}^{-1}\cdot \text{s}^{-2} = \frac{J}{m^{3}}\,,
	\end{equation*}
	which has the dimension of energy density, as desired. 
	
	Next, we analyze the Schwarzschild-like radius $\rho_{\scaleto{S}{4pt}}$ and its respective mass term $m_{\scaleto{S}{4pt}}$. We know that $\rho_{\scaleto{S}{4pt}}/\rho$ is dimensionless because $A(\rho)$ is. Then, it is trivial to conclude that:
	\begin{equation*}
		\left[\rho_{\scaleto{S}{4pt}}\right]_{\scaleto{SI}{4pt}} = \text{m}\,,
	\end{equation*}
	while 
	\begin{equation*}
		\left[m_{\scaleto{S}{4pt}}\right]_{\scaleto{SI}{4pt}} = \left[\frac{G}{c^{2}}\right]^{-1}_{\scaleto{SI}{4pt}}\,\left[\rho_{\scaleto{S}{4pt}}\right]_{\scaleto{SI}{4pt}} = \text{kg}\,,
	\end{equation*}
	according to our aim to treat $m_{\scaleto{S}{4pt}}$ like a mass. On the other hand, the dimension for the quintessence parameter $N_{\scaleto{Q}{4pt}}$ is simply
	\begin{equation*}
		\left[N_{\scaleto{Q}{4pt}}\right]_{\scaleto{SI}{4pt}} = \left[\rho^{2\alpha_{\scaleto{Q}{3pt}}}\right]^{-1}_{\scaleto{SI}{4pt}} = \text{m}^{-2\alpha_{\scaleto{Q}{3pt}}}\,,
	\end{equation*}
	so that $N_{\scaleto{Q}{4pt}}\,\rho^{2\alpha_{\scaleto{Q}{3pt}}}$ is dimensionless.
	
	At last, we give the units associated with an arbitrary scalar field mass, $m_{\phi}$, satisfying the Klein-Gordon equation. We recall that
	\begin{equation*}
		\left[\frac{m_{\phi}^{2}\,c^{2}}{\hbar^{2}}\right]_{\scaleto{SI}{4pt}} = \left[g^{\mu\nu}\,\nabla_{\mu}\nabla_{\nu}\right]_{\scaleto{SI}{4pt}} = \text{m}^{-2}\,,
	\end{equation*}
	implying
	\begin{equation*}
		\left[m_{\phi}\right]_{\scaleto{SI}{4pt}} = \left[\frac{\hbar}{c}\right]_{\scaleto{SI}{4pt}}\times \text{m}^{-1} = \text{kg}\cdot \text{m} \times \text{m}^{-1} = \text{kg}\,.
	\end{equation*}
	Therefore, $m_{\phi}$ is a mass contribution, as it should be. In Sec. \ref{sec-Estimates}, we used the insight from the dimensional analysis to find an admissible range of values for these parameters.

	\section{Confluent Heun equation}\label{app-Heun-equation}
	
	Let us consider the canonical form of the confluent Heun equation, given by \citep{Ronveaux1995,Hortacsu2018}:
	\begin{equation}
		\frac{d^{2} y}{d z^{2}} + \left(\alpha + \frac{\beta + 1}{z} + \frac{\gamma + 1}{z - 1}\right)\frac{d y}{d z} + \left(\frac{\mu}{z} + \frac{\nu}{z - 1}\right)y = 0\,,\label{HeunConfluentCanonical}
	\end{equation}
	with $z={0,\,1}$ being the regular singular points and $\alpha,\,\beta,\,\gamma,\,\mu,\,\nu$ are constants. The irregular singular point lies at infinity. For the case of the standard confluent Heun \emph{function}, it is true that
	\begin{equation}
		\text{HeunC}\left(\alpha,\,\beta,\,\gamma,\,\delta,\,\eta;\,0\right) = 1\,,\label{Initial_Condition_HeunC}
	\end{equation}
	while
	\begin{align}
		\mu &= \frac{1}{2}\left(\alpha - \beta - \gamma + \alpha\beta - \beta\gamma\right) - \eta\,,\label{Mu_Condition}\\
		\nu &= \frac{1}{2}\left(\alpha + \beta + \gamma + \alpha\gamma + \beta\gamma\right) + \delta + \eta\,.\label{Nu_Condition}
	\end{align}
	Writing \cref{HeunConfluentCanonical} in the Liouville normal form \cite{Titchmarsh1962}, we obtain the following differential equation:
	\begin{equation}
		u''(z) + \left[\frac{A}{z} + \frac{B}{(z - 1)} + \frac{C}{z^{2}} + \frac{D}{(z - 1)^{2}} + E\right] u(z) = 0\,,\label{Eq_u_Normal_Form}
	\end{equation}
	where
	\begin{equation}
		\begin{aligned}
			A &= \tfrac{1}{2} - \eta\,,\\
			B &= -\tfrac{1}{2} + \delta + \eta\,,\\
			C &= \left(1 - \beta^{2}\right)/4\,,\\
			D &= \left(1 - \gamma^{2}\right)/4\,,\\
			E &= -\alpha^{2}/4\,.
		\end{aligned} \label{Parameters_HeunC_Normal}
	\end{equation}
	The relation between $u(z)$ and $y(z)$ is determined by   
	\begin{equation}
		u(z) = u_{0}\,\exp\left({\alpha z/2}\right)\,z^{(1+\beta)/2}\,\left(z - 1\right)^{(1 + \gamma)/2}\,y(z)\,,
	\end{equation}
	where $u_{0}$ is a constant. Therefore, any ODE of the form \eqref{Eq_u_Normal_Form} will have the following solution:
	\begin{equation}
		\begin{aligned}
			u(z) &= \exp\left({\alpha z/2}\right)\,z^{(1+\beta)/2}\,\left(z - 1\right)^{(1 + \gamma)/2}
			\left[ c_{1}
			\text{HeunC}\left(\alpha,\,\beta,\,\gamma,\,\delta,\,\eta;\,x\right) \right.\\
			&\left.+ c_{2} \, x^{-\beta}\,
			\text{HeunC}\left(\alpha,\,-\beta,\,\gamma,\,\delta,\,\eta;\,x\right)\right] \,,\label{Sol_Geral_u}
		\end{aligned}
	\end{equation}
	where $c_{1},\,c_{2}$ are arbitrary constants to be set with boundary conditions.

	\subsection{Fuchs-Frobenius solution near z=1}
	
	This section investigates solutions near $z = 1$, which when considering the system presented in \cref{sec-KG-equation} represents a particle close to the event horizon. To do this, consider the confluent Heun equation given by \eqref{HeunConfluentCanonical}. First, we multiply \cref{HeunConfluentCanonical} by a factor of $z(z-1)$ so that
	\begin{equation}
		\begin{aligned}
			0 &= z\left(z-1\right) y''(z)\\ 
			&+ \left[\alpha z\left(z - 1\right) + \left(\beta + 1\right)\left(z - 1\right) + \left(\gamma+1\right) z\right] y'(z)\\ 
			&+ \left[\mu\left(z-1\right) + \nu z\right] y(z)\,.
		\end{aligned} \label{Confluent_Eq_Aberta}
	\end{equation}
	Next, let us define the auxiliary variable $\chi \coloneq z - 1$, such that
	\begin{equation*}
		\frac{d}{d\chi} = \frac{d}{d z}\,.
	\end{equation*}
	In terms of $\chi$, \cref{Confluent_Eq_Aberta} becomes
	\begin{equation}
		\begin{aligned}
			0 &= \chi^{2} \,y'' + \Theta_{c}\,\chi \,y' + \nu \,y\\
			&+ \chi \,y'' + \left(\gamma + 1\right) y'\\
			&+ \alpha \,\chi^{2} \,y' + \left(\mu + \nu\right) \chi \,y\,,
		\end{aligned} \label{Eq_para_chi}
	\end{equation}
	where 
	\begin{equation*}
		\Theta_{c} \coloneq \alpha + \beta + \gamma + 2\,,
	\end{equation*}
	while the prime notation ($\,'\,$) means a derivative with respect to $\chi$, i.e. $d/d \chi$.
	
	In the sequence, we plug the ansatz
	\begin{equation}
		y(\chi) = \sum_{m=0}^{+\infty} c_{m}\,\chi^{m+r}\,,\label{series-chi}
	\end{equation}
	into \cref{Eq_para_chi} to show that
	\begin{equation}
		\begin{aligned}
			0 &= \sum_{m=0}^{+\infty} c_{m} \left[\left(m+r\right)\left(m+r+\Theta_{c}-1\right) + \nu\right] \chi^{m+r}\\
			&+ \sum_{m=0}^{+\infty} c_{m} \left(m+r\right)\left(m+r+\gamma\right) \chi^{m+r-1}\\
			&+\sum_{m=0}^{+\infty} c_{m} \left[\alpha\left(m+r\right) + \left(\mu + \nu\right)\right] \chi^{m+r+1}\,.
		\end{aligned}\label{Confluent_z_1_case}
	\end{equation}
	Now, we rename the summation index $m$ in \cref{Confluent_z_1_case} to obtain that
	\begin{equation}
		\begin{aligned}
			0 &= \sum_{k=0}^{+\infty} c_{k} \left[\left(k+r\right)\left(k+r+\Theta_{c}-1\right) + \nu\right] \chi^{k+r}\\
			&+ \sum_{k=-1}^{+\infty} c_{k+1} \left(k+r+1\right)\left(k+r+\gamma+1\right) \chi^{k+r}\\
			&+\sum_{k=1}^{+\infty} c_{k-1} \left[\alpha\left(k+r-1\right) + \left(\mu + \nu\right)\right] \chi^{k+r}\,.
		\end{aligned}\label{Confluent_Expanded_Around_1}
	\end{equation}
	It follows from \cref{Confluent_Expanded_Around_1} that:
	\begin{subequations}
		\begin{align}
			&c_{0} \,r \left(r + \gamma\right) = 0\,,\label{Equação_Indicial}\\
			&c_{1} = \frac{r\left(r + \alpha + \beta + \gamma + 1\right) + \nu}{\left(r+1\right)\left(r+\gamma+1\right)} \,c_{0}\,, \label{c_1_Confluente_1}
		\end{align}
		while we write the general term as
		\begin{equation}
			M_{0} \,c_{m} + M_{1} \,c_{m+1} + M_{2} \,c_{m+2} = 0\,, \quad m\geq 0\,,\label{Termo_Geral}
		\end{equation}
		with
		\begin{align}
			M_{0} &= \alpha\left(m+r\right) + \left(\mu+\nu\right)\,,\label{M_0}\\
			M_{1} &= \left(m+r+1\right)\left(m+r+\alpha+\beta+\gamma+2\right) + \nu\,,\label{M_1}\\
			M_{2} &= \left(m+r+2\right)\left(m+r+\gamma+2\right)\,.\label{M_2}
		\end{align}
	\end{subequations}
	Note that we wrote the final result in terms of $m = k-1$ for convenience. 
	Finally, we can set $c_{0} = 1$ to obtain explicitly the first three coefficients of the local solution to the confluent Heun equation near $z=1$, given by
	\begin{equation}
		\text{HeunC}\ell\left(\alpha,\,\beta,\,\gamma,\,\delta,\,\eta;\,z-1\right) = 1 + c_{1} \left(z-1\right) + c_{2} \left(z-1\right)^{2} + \dots\,,\label{HeunCl_around_1}
	\end{equation}
	where
	\begin{equation}
		c_{1} = \frac{r\left(r + \alpha + \beta + \gamma + 1\right)}{\left(r+1\right)\left(r+\gamma+1\right)} + \frac{\left(\alpha + \beta\right)\left(1+\gamma\right)+ \gamma + 2 \left(\delta+\eta\right)}{2 \left(r+1\right)\left(r+\gamma+1\right)}\,,
	\end{equation}
	and
	\begin{align}
		c_{2} &= -\frac{1}{\left(r + 2\right)\left(r + \gamma + 2\right)}\left\{\frac{\alpha}{2}\left(2r+\beta+\gamma+2\right) + \delta \right.\notag\\
		&+\left. c_{1}\left[\left(r+1\right)\left(r+\alpha+\beta+\gamma+2\right) + \frac{\gamma}{2} + \delta + \eta +\frac{1}{2}\left(\alpha+\beta\right)\left(1+\gamma\right)\right] \right\}\,.
	\end{align}
	
	\subsubsection{Imposing the break off of the series solution - polynomial solutions}
	
	According to the solution given by \cref{Equação_Indicial,c_1_Confluente_1,Termo_Geral,M_0,M_1,M_2}, one can demand the break off of the series \eqref{series-chi} into a polynomial requiring that $M_{0} = 0$ and $c_{m+1}\left(\alpha,\,\beta,\,\gamma,\,\delta,\,\eta\right) = 0$ for some $m = N$.
	
	The first requirement, i.e. $M_{0} = 0$ for some $m=N\in\mathbb{Z}^{+}$ gives us the condition
	\begin{equation}
		\alpha \left(N+1+\frac{\beta+\gamma}{2}\right) + \delta = 0\,,
	\end{equation}
	which can be used to determine the energy levels in terms of $N$.
	
	\bibliographystyle{elsarticle-num} 
	\bibliography{References.bib}

\end{document}